\documentclass[usenatbib,iop,numberedappendix]{aeb_emulateapj_2010}
\usepackage{amsmath}
\usepackage{amssymb}
\usepackage{xspace}
\usepackage[normalem]{ulem}

\usepackage{natbib}

\usepackage[usenames,dvips]{color}

\def\s{{\rm s}}

\def\yr{{\rm yr}} 
\def\Gyr{{\rm G}\yr}

\def\m{{\rm m}}

\def\mum{\mu\m} 
 
\def\cm{{\rm c}\m} 
\def\km{{\rm k}\m} 

\def\pc{{\rm pc}} 
 
\def\Mpc{{\rm M}\pc} 
\def\Gpc{{\rm G}\pc}

\def\eV{{\rm eV}} 
\def\keV{{\rm k}\eV} 
\def\MeV{{\rm M}\eV} 
\def\GeV{{\rm G}\eV} 
\def\TeV{{\rm T}\eV}

\def\erg{{\rm erg}} 
 
\def\K{{\rm K}} 
\def\Ry{{\rm Ry}}

\def\Sr{{\rm sr}}

\def\del#1{{}}

\newcommand{\rmn}{\mathrm}
\newcommand{\e}{{\rm e}}

\def\GIC{\Gamma_{\rm IC}}

\def\GM{\Gamma_{\rm M,k}}

\def\Dpp{D_{\rm pp}}
\def\bDpp{\bar{D}_{\rm pp}}

\def\nb{n_{\rm b}}
\def\nIGM{n_{\rm IGM}}
\def\nbaryon{n_{\rm bary}}
\def\Qcorr{\dot{\mathcal{Q}}_{\rm corr}}

\def\Lya{Ly$\alpha$\xspace}

\def\LIR{L_{\rm IR}}

\def\Fermi{{\em Fermi}\xspace}

\def\QLF{\phi_Q}
\def\BLF{\phi_B}
\def\tQLF{\tilde{\phi}_Q}
\def\tBLF{\tilde{\phi}_B}

\def\dlL{d\!\log_{10}L}

\def\dq{\dot{Q}}

\begin{document}

\title{
The Cosmological Impact of Luminous TeV Blazars II:\\
Rewriting the Thermal History of the Intergalactic Medium
}

\author{
Philip Chang\altaffilmark{1,2},
Avery E.~Broderick\altaffilmark{1,3,4},
and
Christoph Pfrommer\altaffilmark{5,1}
}
\altaffiltext{1}{Canadian Institute for Theoretical Astrophysics, 60 St.~George Street, Toronto, ON M5S 3H8, Canada; aeb@cita.utoronto.ca, pchang@cita.utoronto.ca}
\altaffiltext{2}{Department of Physics, University of Wisconsin-Milwaukee, 1900 E. Kenwood Boulevard, Milwaukee, WI 53211, USA}
\altaffiltext{3}{Perimeter Institute for Theoretical Physics, 31 Caroline Street North, Waterloo, ON, N2L 2Y5, Canada}
\altaffiltext{4}{Department of Physics and Astronomy, University of Waterloo, 200 University Avenue West, Waterloo, ON, N2L 3G1, Canada}
\altaffiltext{5}{Heidelberg Institute for Theoretical Studies, Schloss-Wolfsbrunnenweg 35, D-69118 Heidelberg, Germany; christoph.pfrommer@h-its.org}

\shorttitle{The Cosmological Impact of Blazar TeV Emission II}
\shortauthors{Chang, Broderick, \& Pfrommer}

\begin{abstract}
The Universe is opaque to extragalactic very high-energy gamma rays (VHEGRs, $E>100\,\GeV$) because they  annihilate and pair produce on the extragalactic background light. 
The resulting ultra-relativistic pairs are commonly assumed to lose
  energy primarily through inverse Compton scattering of cosmic microwave
  background photons, reprocessing the original emission from TeV to GeV
  energies.
  In \citet[][Paper I of this three paper series]{BCP},
  we argued that this is not the case; powerful
  plasma instabilities driven by the highly anisotropic nature of the
  ultra-relativistic pair distribution provide a plausible way to
  dissipate the kinetic energy of the TeV-generated pairs locally,
  heating the intergalactic medium (IGM).
Here, we explore the effect of this heating upon the thermal history of the
IGM.
We collate the observed extragalactic VHEGR sources to determine a local
VHEGR heating rate.
Given the pointed nature of VHEGR observations, we estimate the
correction for the various selection effects using \Fermi observations
of high and intermediate peaked BL Lacs.
As the extragalactic component of the local VHEGR flux is dominated by
TeV blazars, we then estimate the evolution of the TeV blazar
luminosity density by tying it to the well-observed quasar luminosity
density, and producing a VHEGR heating rate as a function of redshift. This heating is relatively homogeneous for $z
\lesssim 4$, but there is greater spatial variation at higher redshift (order unity at $z\sim 6$) because of the reduced number of blazars that contribute to local heating.    
We show that this new heating process dominates photoheating in the
low-redshift evolution of the IGM and calculate the effect of this
heating in a one-zone model.
As a consequence, the inclusion of TeV blazar heating qualitatively
and quantitatively changes the structure and history of the IGM.
Due to the homogeneous nature of the extragalactic background light,
TeV blazars produce a uniform {\em volumetric} heating rate.
This heating is sufficient to
increase the temperature of the mean density IGM by nearly an order of
magnitude, and at low densities by substantially more.  It also naturally produces the inverted temperature-density relation inferred by
recent observations of the high-redshift \Lya forest, a feature that
is difficult to reconcile with standard reionization models.
Finally, we close with a discussion on the possibility
of detecting this hot low-density IGM suggested by our model either
directly or indirectly via the local \Lya forest, the Comptonized cosmic microwave background, or free-free emission, but find that such
measurements are currently not feasible. 
\end{abstract}


\maketitle

\section{Introduction} \label{I}

The \Fermi satellite and ground based imaging atmospheric Cerenkov telescopes such as
H.E.S.S., MAGIC, and VERITAS\footnote{High Energy
  Stereoscopic System, Major Atmospheric Gamma Imaging Cerenkov Telescope, Very
  Energetic Radiation Imaging Telescope Array System.} have demonstrated that the ultra-high
energy Universe is teeming with energetic very high-energy gamma-ray 
(VHEGR, $E > 100\,\GeV$) sources, the extragalactic component of which mainly consists of TeV
blazars with a minority population of other sources  such as radio galaxies
and starburst galaxies.  These VHEGR observations are being used
to constrain the sites and mechanisms 
of particle acceleration
\citep[see, e.g., ][]{Pagl_etal:96,Domi-Torr:05,Thom-Quat-Waxm:07,Pers-Reph-Arie:08,deCe-Torr-Rodr:09,Reph-Arie-Pers:10,Lack_etal:10},
dynamics of black hole jets
\citep[see, e.g., ][]{Jones+1974,Ghisellini+1989,Ghisellini+2008,Tavecchio+2008,Ghisellini+2009},
and intergalactic magnetic fields
\citep[IGMF; ][]{Nero-Vovk:10,Tave_etal:10a,Tave_etal:10b,Derm_etal:10,Tayl-Vovk-Nero:11,Dola_etal:11,Taka_etal:11,Vovk+12}.

While these objects have an interesting phenomenology, they are believed to have 
a minor impact upon the Universe at large,
i.e., on the formation of structures and thermodynamics.  Energetically this is not an unreasonable assumption, as VHEGR emission is $\sim 0.1$\% of the radiative power of quasars.   
However, in this series of papers, we show that in spite of their energetic disadvantage, TeV blazars have a significant effect on structure formation and a dominant effect on thermodynamics -- that is the VHEGR emission
from blazars ``punches'' far above its energetic weight.
Namely, {\em if} the radiation from VHEGR sources is
thermalized, as we have argued is the case in \citet[hereafter Paper I]{BCP},
the heating due to these VHEGRs
dominates photoionization heating throughout the vast majority of the
Universe at $z\lesssim 3$, raising the temperature of the low-density
IGM by up to two orders of magnitude.

Given that the total power emitted by AGNs and stars in the UV and X-rays
vastly exceeds that due to the TeV blazars, it seems counterintuitive
that blazar heating dominates photoheating.  However, the UV and
X-ray background heats the intergalactic medium (IGM) inefficiently
after reionization, while VHEGR photons heats the IGM efficiently via plasma beam
instabilities (Paper I).  This difference in the heating efficiency of
photoheating vs. blazar heating is due to the difference in the
rate-limiting process in each.  The rate of photoheating after
reionization is not limited by the availability of ionizing photons,
but instead by abundance of targets, and thus recombination. On the
other hand, the heating of the IGM by TeV blazars does not suffer from
a deficit of targets and is only limited by the total cosmic power of
VHEGR sources.  

We illustrate this point with the following order of magnitude
estimate.  First let us estimate the amount of photoheating that the
IGM suffers at the present day.  The recombination rate of H is of
order the Hubble time at the mean density of the Universe at present.
Hence, an average H atom in the IGM will recombine once over a Hubble
time only to be ionized immediately by a UV photon.  Using a spectral
index of -1.6 for the ionizing background, which is appropriate for
quasars \citep{Furlanetto08}, the average amount of excess energy
absorbed per ionization is $\epsilon_{\rm exc} \approx 4\,{\rm eV}$.
The fraction of the rest mass energy of all the baryons in the
universe required to produce this amount of heating in the IGM is
\begin{equation}\label{eq:excess}
f_{\rm exc} = \frac {\epsilon_{\rm exc}}{m_p c^2} \approx 4\times 10^{-9}\left(\frac{\epsilon_{\rm exc}}{4\,{\rm eV}}\right).
\end{equation}
Hence, only a small amount of rest mass energy is injected as thermal energy
into the IGM. An equivalent statement is that the diffuse IGM is optically thin
to ionization radiation.\footnote{By ``diffuse'' IGM, we discount Lyman limit
  systems.}

By comparison, the fraction of the
baryon rest mass locked up in massive black holes in the Universe is
$f_{\rm BH} \approx 9\times 10^{-5}$ and in stars is $0.06$
\citep{Fukugita+04}. Assuming a radiative efficiency (relative to rest
mass) for black holes and stars of $0.1$ and $10^{-3}$ respectively,
we find that fraction of the baryon rest mass converted to radiation
in black holes and stars is $\epsilon_{\rm rad,BH} \sim 10^{-5}$ and
$\epsilon_{\rm rad,*}\sim 6\times 10^{-5}$, respectively.\footnote{
  Here we have likely significantly overestimated the radiative
  efficiency of the stellar component for two reasons.  First, most of
  the mass locked up in the stellar component is contained in low-mass
  stars, which are capable of converting only a small fraction of
  their rest mass into energy.  Second, we have not accounted for the
  fraction of stellar radiation that is capable of ionizing the IGM.}
This is many orders of
magnitude larger than what is required by Equation
(\ref{eq:excess}) and demonstrates an important point: {\it hard
  ionizing radiation is inefficient at heating the IGM after H and He
  reionization.} 

In Paper I, we argued that VHEGR photons are efficiently converted into heat in the IGM via plasma instabilities.  Hence, the limiting factor is the total energy density of VHEGR photons over cosmic time.  To estimate this energy density, we note that the local {\it observed} TeV blazar luminosity density is $2.1\times 10^{-3}$ that of the local quasar luminosity density (see Section \ref{sec:z=0 heating}), after correcting for various selection effects. Assuming that the TeV blazar luminosity density tracks the quasar luminosity density over cosmic time, this implies that the fraction of baryon rest mass that is converted to VHEGR photons (and ultimately heating of the IGM) is
\begin{equation}
f_{\rm TeV} = \frac{\textrm{TeV Blazar Luminosity Density}}{\textrm{Quasar Luminosity Density}} \times \epsilon_{\rm rad,BH} = 2.1\times 10^{-8},
 \end{equation}
which is nearly five times that due to photoheating (i.e., Equation (\ref{eq:excess})) and demonstrates the dominance of TeV blazar heating. Equivalently, the greater efficiency of TeV blazar heating compared to photoheating more than makes up for its energetic disadvantage.

The physics of this heating and its cosmological consequences is
the subject of this series of three papers.  
In Paper I, we studied the physics of VHEGR photon propagation through
the Universe.  As these VHEGRs propagate through the Universe, they interact
with the soft photons that comprise the 
extragalactic background light (EBL) and produce ultra-relativistic
pairs \citep[see, e.g., ][]{Goul-Schr:67,Sala-Stec:98,Nero-Semi:09}.  Typical mean free paths are between $30\,\Mpc$ and $1\,\Gpc$,
depending upon the energy of the VHEGR and redshift, i.e., the Universe is optically thick to VHEGR.  The result is a ubiquitous population of ultra-relativistic pairs, with typical Lorentz factors of $10^5$--$10^7$. Previously, it has been assumed that they lose energy exclusively
through inverse-Compton scattering the cosmic microwave background
(CMB) and EBL, producing GeV gamma rays that form part of the EGRB \citep[see, e.g., ][]{Naru-Tota:06,Knei-Mann:08,Inou-Tota:09,Vent:10}. The non-observation of this GeV gamma-rays has been used to argue for cosmologically interesting IGMFs
\citep{Nero-Vovk:10,Tave_etal:10a,Tave_etal:10b,Derm_etal:10,Tayl-Vovk-Nero:11,Dola_etal:11,Taka_etal:11,Vovk+12}.

We then presented a plausible alternative mechanism for extracting
the kinetic energy of the ultra-relativistic pairs: plasma beam instabilities.
Despite the extraordinarily
dilute nature of this ultra-relativistic pair plasma, we found a variety of plasma instabilities which grow on timescales
short in comparison to the inverse-Compton cooling time, the most
important of which is the ``oblique'' instability 
\citep{Bret-Firp-Deut:04}.  Via this instability, these ultra-relativistic pairs lose their kinetic energy by depositing it as heat in the IGM.  
Because these beams then cool well before an inverse-Compton cascade (ICC) can develop, the simplest versions
of the argument used to produce limits upon the IGMF are precluded.
Hence, the existence of the IGMF does not follow from the non-observation of GeV gamma rays from existing TeV
blazars as previous groups have argued.  In
addition, the lack of an ICC allows for a large
and evolving blazar population without upsetting the \Fermi limits on
the EGRB and statistics of high-energy blazars.  

Based upon the plasma-instability mechanism, we now adopt as a hypothesis
that the ultra-relativistic pairs primarily deposit their energy in
the IGM via this or a related mechanism.
In this paper (Paper II), we explore the impact of this heating on the
thermodynamics of the IGM.  We estimate the amount of heating provided
by the observed TeV blazar population after correcting for the
selection effects of the current pencil beam VHEGR observations using
the all-sky monitoring of the \Fermi satellite.  We will show that the
luminosity density in VHEGRs is of order $0.2\%$ of the quasar
luminosity density, which dominates the
photoheating rate at low $z$. We then explore the qualitative and
quantitative nature of this heating, which in essence, serves as an
alternate feedback mechanism.   In particular, TeV blazars
deposit heat evenly in a {\em volumetric} sense, i.e., independent of
the local IGM density.  Hence, this heating deposits more energy per
baryon in low-density regions than in high-density regions,
naturally producing an inverted temperature-density relation in voids.  With only a
minor rescaling of the empirically normalized of observed blazar
heating, we find that it is possible to reproduce the inferred
inverted temperature-density relation at $z=2-3$ \citep{Bolton+09,Viel+09},
something which has proven to be a problem within the context of
standard reionization models \citep{McQuinn+09,Bolton+09}.

In \citet[][hereafter Paper III]{PCB}, we will explore the result of this additional IGM heating upon
the formation of structure in the Universe.  In particular, we will show that
the injection of entropy into the IGM by TeV blazars contributes to
developing a redshift dependent entropy floor for galaxy clusters and groups at
$z\lesssim 2$ and suppresses the formation of dwarfs.  We will highlight
that the redshift dependent nature of TeV blazar heating in our model
suggests a large injection of entropy around $z\sim 1$, which boosts
the entropy of late forming objects.  This predicted enhanced entropy
of young groups is consistent with recent observations 
that show
optically bright and therefore young, groups and
clusters are X-ray dim -- that is having a lower gas density due to a
raised entropy floor.  We also will show that TeV
blazar heating suppresses the formation of late forming dwarfs both in
galactic halos, i.e., the missing satellite problem \citep{Krav:10}, and
in voids, i.e., the void phenomenon \citep{Peebles2001} by raising the
temperature of the IGM such that gas cannot collapse to form galaxies.

This work is organized as follows:
We first review the fate of energy carried by VHEGR photons in Section
\ref{sec:pair propagation}, discussed in detail in Paper I. We describe how VHEGR photons produce pairs in the IGM and how
these ultra-relativistic pair beams are unstable to plasma
instabilities.  In particular, we highlight the ``oblique''
instability, which is especially efficient at converting the kinetic
energy of the beams into thermal energy in the IGM.  Motivated by our review
of Paper I, we adopt the assumption that the kinetic energy of
the ultra-relativistic pairs is thermalized in the IGM, either via the
``oblique'' instability or some related mechanism.
In Section \ref{sec:IGMHR}, we 
estimate the current TeV-blazar IGM heating rate 
by collating the known extragalactic TeV blazars with a well measured spectrum,
accounting for incompleteness (Section \ref{sec:z=0 heating}).
We use the similarity between the luminosity functions of nearby
quasars and TeV blazars found in Paper I, to extend the heating rate
to $z>0$ and estimate the TeV-blazar covering fraction (Sections
\ref{sec:z>0 heating} and \ref{sec:homogeneity}).
The implications for the thermal history of low-density regions (less
than 10 times the mean density, i.e., $1+\delta\lesssim10$) are
explored in Section \ref{sec:thermal history}.  Generally, we find
that without any fine tuning it is possible to reproduce the inverted
temperature-density relation at $z=2-3$ inferred by high-redshift \Lya studies
\citep{Bolton+08,Viel+09}, while simultaneously satisfying the
temperature constraints at $z=2$ \citep[e.g., those by ][]{Lidz+10}
and leaving the local \Lya forest unaffected.

The results of this work and Paper III assume that the energy of TeV
blazars are efficiently thermalized in the IGM.  Though we have
identified a particularly promising instability, i.e., the "oblique"
instability, in Paper I, the particular details by which the energy in
ultra-relativistic pairs is thermalized are unimportant.
Instead, the results of this work and Paper III depend solely on the 
gross energetics of the TeV blazars.  As a consequence, observations
of the thermal history and constraints upon the structures in
low-density regions represent an independent empirical probe of the
fate of the ultra-relativistic pairs produced by TeV blazars.

For all of the calculations presented below (and in this series) we have assumed the WMAP7
cosmology with $h_0 = 0.704$, $\Omega_{DM} = 0.227$,
$\Omega_{B} = 0.0456$, and $\Omega_{\Lambda} = 0.728$ \citep{WMAP7_2011}.\\

\section{Review of VHEGR Photon Propagation through the IGM}\label{sec:pair propagation}

The observed extragalactic VHEGR sources are all located at low redshift
($z\lesssim0.5$).  This is a due to the annihilation of
VHEGRs on the EBL, producing ultra-relativistic
$e^\pm$ pairs \citep[see, e.g., ][]{Goul-Schr:67,Sala-Stec:98,Nero-Semi:09}.
Here we describe the fate of these VHEGRs, and the consequences of
propagating through the intergalactic medium (IGM) for the pairs they
produce. We refer the interested reader to Sections 2 and 3 of Paper I for a more complete
discussion of the properties of the generated pairs and the importance
and nature of plasma beam instabilities upon their propagation.

VHEGRs are attenuated by
the pair production off EBL photons.  Namely, when
the energies of the VHEGRs ($E$) and the EBL photon ($E_{\rm EBL}$)
exceed the rest mass energy of the $e^\pm$ pair in the center of mass
frame, i.e., $2E\,E_{\rm EBL}(1-\cos\theta) > 4m_e^2 c^4$, where $\theta$ is the
relative angle of propagation in the lab frame,
an $e^\pm$ pair can be produced with Lorentz factor
$\gamma\simeq E/2m_e c^2$ \citep{Goul-Schr:67}.
An estimate for the mean free path of VHEGR photons is (Paper I)
\begin{equation}
\Dpp(E,z)
=
35\left(\frac{E}{1\,\TeV}\right)^{-1}
\left(\frac{1+z}{2}\right)^{-\zeta}\,\Mpc\,,
\label{eq:Dpp}
\end{equation}
where the redshift evolution is due to that of the EBL alone, is dependent
predominately upon the star formation history, and $\zeta=4.5$ for
$z<1$ and $\zeta=0$ for $z\ge1$
\citep{Knei_etal:04,Nero-Semi:09}\footnote{Despite the fact that the
  EBL contribution from starbursts peaks at $z=3$ and declines rapidly
  afterward, galaxies and Type 1 AGNs compensate for the lost flux
  until $z=1$.  See, for example, Figure 3 from
  \citet{Fran-Rodi-Vacc:08}.}.
Relative to the Hubble length, the attenuation length of VHEGRs is very short. In fact, above $100\,\GeV$
the Universe is optically thick to sources at $z>1$
\citep[cf.][]{Fran-Rodi-Vacc:08}.

Since $\Dpp$ is much larger than any conceivable source size, and
$E_{\rm EBL}\ll E$, locally these pairs necessarily constitute a cold,
highly anisotropic beam.
The production of pairs by the interaction of VHEGRs
with the EBL is opposed by the removal of these pairs by various
cooling processes. 
That is, the evolution of the characteristic pair-beam density at
the injection Lorentz factor, $\nb$, is governed by the Boltzmann equation:
\begin{equation}
\frac{\partial \nb}{\partial t}
+
\frac{c}{r^2} \frac{\partial r^2 \nb}{\partial r}
+
\Gamma \nb = \dot{\nb},
\end{equation}
where the left-hand side assumes all the pairs are moving away from
the VHEGR source relativistically ($v^r=c$ and $p^r=\gamma m_e c$), 
the right-hand side corresponds to pair production, and $\Gamma$ is
the cooling/removal rate of these pairs. 
In a homogeneous steady state, the rate of production is balanced by
the rate of removal, which gives:
\begin{equation}
\nb \simeq \frac{2 F_E}{\Dpp \Gamma}\,,
\label{eq:nb}
\end{equation}
where the rate of production is given by
$\dot{n}_{\rm beam} = 2 (E dN/dE)/\Dpp = 2 F_E/\Dpp$.
Generally, $\Gamma$ is a function of energy and beam
energy and refer the reader to Paper I for additional
details.
However, for any choice of $\Gamma$, the solution to Equation
(\ref{eq:nb}) gives $\nb(E,F_E,z)$.

Commonly, it is assumed that the pairs evolve primarily due to
an
ICC which deposits the energy from the original
VHEGR photon near $\sim100\,\GeV$. 
Note that this is fundamentally radiative; after the VHEGRs are
scattered down to $100\,\GeV$ they effectively decouple from the
Universe.
ICCs of these pair beams widely exploited as a
possible probe of IGM magnetic fields in the context of the missing
inverse Compton features at $100\,\GeV$
\citep[see, e.g.,][]{Nero-Vovk:10,Tave_etal:10a,Tave_etal:10b,Derm_etal:10,Tayl-Vovk-Nero:11,Taka_etal:11,Dola_etal:11}.
The associated cooling rate for this process for pairs with Lorentz factor $\gamma$ is
\begin{equation}\label{eq:IC}
\GIC=
\frac{4\sigma_T u_{\rm CMB}}{3 m_e c} \gamma
\simeq
1.4\times10^{-20}(1+z)^4\gamma \,\,\s^{-1}\,,
\end{equation}
where $\sigma_T$ denotes the Thompson cross section.
The strong redshift dependence arises from the rapid increase in the
CMB energy density with $z$ ($u_{\rm CMB}\propto(1+z)^4$).

However, this may not be the dominant process.
As we showed in Paper I, plasma beam instabilities are a potential
mechanism by which the kinetic energy of the pairs is extracted much more
rapidly.
The two plasma beam instabilities most frequently discussed, the
two stream and Weibel instabilities, are particular limits of a more
general instability, differentiated by the direction of the perturbed
wave vector with respect to the beam orientation \citep[perpendicular for
Weibel, parallel for the two stream;][]{Bret-Firp-Deut:05}.
For dilute beams, by far the most powerful growth occurs at oblique
angles, and thus named the ``oblique'' instability.  Paper I and 
\citet{Bret-Grem-Diec:10} gives a more extensive discussion of these
three plasma instabilities.

Relative to the laboratory frame, these pair beams are ``cold'', i.e.,
their transverse momentum is much smaller than their parallel
momentum, but even the small transverse temperatures that are acquired
from pair production for these pair
beams are important, placing the oblique instability in the
  kinetic regime.  Here, the oblique instability cooling rate has
been numerically measured to be
\begin{equation}
\GM
\simeq
0.4 \frac{m_e c^2}{k T_b} \frac{\nb}{\nIGM} \omega_P
\simeq
0.4 \gamma \sqrt{\frac{4\pi e^2 \nb^2}{m_e \nIGM}}\,,
\end{equation}
where $\nIGM = 2.2\times10^{-7}(1+\delta)(1+z)^3\,\cm^{-3}$ is
the IGM free-electron number density (assuming full ionization), $e$
is the elementary charge, and we have set the beam temperature in the
beam frame to $m_e c^2/k$, characteristic of that induced by pair
production, and thus $kT_b=m_e c^2/\gamma$ \citep{Bret-Grem-Beni:10}. Both
the cold and hot growth rates have been verified explicitly using 
particle-in-cell (PIC) simulations, though at somewhat less dilute
beams than the pair beams from TeV blazars \citep{Bret-Grem-Diec:10}.

For the kinetic oblique instability, the cooling rate is 
\begin{multline}
\GM
\simeq
3.6\times10^{-11} 
\left(1+\delta\right)^{-1/4}
\left(\frac{1+z}{2}\right)^{(6\zeta-3)/4}\\
\times\left(\frac{E L_E}{10^{45}\,\erg\,\s^{-1}}\right)^{1/2}
\left(\frac{E}{\TeV}\right)^{3/2}
\s^{-1}\,,
\end{multline}
where  $L_E$ is the isotropic-equivalent luminosity per unit energy of the VHEGR source.
This is a stronger function of photon energy than inverse-Compton
cooling, implying that it will eventually dominate at sufficiently
high energies, assuming a flat VHEGR spectrum.  In addition it is a very
weak function of $\delta$, being only marginally faster in
lower-density regions, and thus the cooling of the pairs is largely
independent of the properties of the background IGM.

The effective cooling rates induced by the "oblique" instability can be
obtained by numerically solving Equation (\ref{eq:nb}) for $\nb$, with
$\Gamma=\GIC+\GM$ as described in Paper I.  For the purpose of this
paper, for TeV blazars the effective cooling rate is generally
dominated by the "oblique" mode and thus the fraction of VHEGR emission
that is effectively thermalized is nearly unity. 
Therefore, we are justified in assuming that all of the VHEGR emission
is thermalized.  In Section \ref{sec:z>0 heating} we will briefly
revisit this assumption at high $z$, finding again that it is well
justified at all $z$.\footnote{We, however, have made the implicit assumption that the nonlinear state of the "oblique" instability removes kinetic energy from the beam at the growth rate of the linear instability.}

However, this may not be the case for all potential VHEGR sources.  
The physics of the "oblique" instability, like all plasma beam
instabilities, depends strongly upon the beam density.
For the pair beams that result from VHEGR-EBL photon interactions, the
beam density is a function of source luminosity.  Hence for 
sufficiently low luminosity systems, the "oblique" mode grows so
slowly that inverse Compton off of the CMB dominates it.  
This defines
a critical isotropic-equivalent luminosity for plasma beam
instabilities to be relevant, typically near $10^{42}\,\erg\,\s^{-1}$,
though depending upon redshift (see Figure 3 of Paper I).
For the TeV blazars considered in this paper, this is generally not a
concern.  However, this will be important in our discussion of
alternative VHEGR sources in Section \ref{sec:alternatives}.

\section{TeV Blazar Heating of the IGM} \label{sec:IGMHR}

As we have argued in Paper I and reviewed in Section \ref{sec:pair
  propagation},  plasma instabilities on the pair beam dissipate the
bulk of the VHEGR emission.  In particular, the
"oblique" instability appears to be a promising mechanism by which the
ultra-relativistic pair beams that are produced from the interaction
of VHEGR and EBL photons blazars are efficiently thermalized and
converted to local IGM heating.  Henceforth, we will make the
assumption for the rest of this work and again in Paper III  that the energy
of TeV blazars is efficiently thermalized in the IGM. With this
stated assumption, we now discuss the sources 
that dominate the IGM heating rate, estimate the magnitude of
the heating rate and its evolution with redshift, and
briefly assess the homogeneity with which this new process occurs.

\subsection{TeV-Blazar Heating Rates} \label{sec:TBHR}
The local heating rate associated with the dissipation of the
high-energy emission from a single blazar is determined by two
factors: the rate at which the VHEGRs are converted
into pairs and the rate at which the energy of the pairs is subsequently
converted locally into heat.  The former is determined by the
pair-production cross-section.  Due to the efficiency of the plasma
instabilities in dissipating the beam energy, the latter is set by the
ratio of the relevant cooling rates.  That is, the single-blazar
heating rate is
\begin{equation}
\dot{q} = \int dE \frac{\theta(E)}{\Dpp(E,z)} f(F_E,E,z) F_E\,,
\label{eq:qdot}
\end{equation}
where $\theta(E)$ is a dimensionless function due to the pair-creation
threshold and which depends upon the shape of the EBL spectrum (we set
$\theta(E)$ to vanish for $E<100\,\GeV$ and be unity otherwise), and
\begin{equation}
f(F_E,E,z) = 1-f_{\rm IC} = \frac{\GM}{\GIC+\GM}\,,
\end{equation}
which is a function of $F_E$, $E$, and $z$ via the dependence of the
cooling rates upon $\nb$, $\gamma$, and $z$.  Generically, $f$ represents the fraction of pair energy that is thermalized, but we have chosen a specific form of $f$ corresponding to the "oblique" instability to make the discussion below more concrete.  While it is also very
weakly dependent upon $\delta$, this may be neglected in practice.

Within the linear regime the plasma instabilities responsible for the
dissipation of pair beam energy are independent of those arising
from beams in substantially different directions.  Since
the local VHEGR flux is dominated by a number of sources that is small in comparison to that needed to isotropize the beam's phase space,
we may treat the resulting evolutions of their associated pair beams
independently and sum their resulting heating rates to determine the
total heating rate, $\dq$.  \footnote{We should note that the if the
  number of blazars in the sky is so numerous as to make the phase
  space distribution of beam particles isotropic, certain classes of
  plasma instabilities -- in particular the Weibel instability -- will
  be suppressed. The volume filling factor of a pair beam in phase
  space is the ratio of perpendicular beam temperature to the beam
  energy, $kT_\rmn{b}/(m_e c^2) \sim 10^{-6}$ (see Paper
  I); hence we would need $\mathcal{N}\sim 10^{12}$ blazars to
  isotropize any given point in space.  However, as will be shown in
  Section \ref{sec:homogeneity} the number of blazars that contribute
  substantially to the local heating rate is much smaller; hence we
  believe the local pair distribution function is sufficiently
  anisotropic.}

\subsubsection{Estimating the Local Heating Rate}\label{sec:z=0 heating}

The local heating rate can be estimated in at least two ways, both of
which give similar results.  First, since the high-energy gamma rays
deposit their energy locally, we can identify the local heating rate
with the high-energy gamma-ray luminosity density of TeV blazars.
In Paper I we showed that this is roughly $2.1\times10^{-3}$ that of
quasars, and thus approximately 
$\sim(0.5$--$1.4)\times10^{38}\,\erg\,\Mpc^{-3}\,\s^{-1}=(3.4$--$9.7)\times10^{-8}\,\eV\,\cm^{-3}\,\Gyr^{-1}$,
depending upon the minimum luminosity at which the heating mechanism
operates.

Alternatively, given a sufficiently complete sample, we can estimate
the local heating rate using that implied by the fluxes of the
observed TeV blazars.  In the present epoch, $f(F_E,E,z)\simeq 1$ at
the relevant $E$, and thus the heating rate takes the particularly
simple form:
\begin{equation}
\left.\dq\right|_{z=0}\simeq\sum_{\rm AGN} \int dE \theta \frac{F_{E,i}}{\Dpp}\,.
\label{eq:qdotest2}
\end{equation}
To evaluate this sum, we have collated the presently 46 extragalactic VHEGR
sources known.\footnote{See http://www.mppmu.mpg.de/$\sim$rwagner/sources/ for an
  up-to-date list.}  Of these 46, only 28 have published measurements of their
VHEGR flux from a combination of VERITAS, H.E.S.S., and MAGIC observations.  For
these 28 sources, we have extracted the parametrized spectra assuming the form,
\begin{equation}\label{eq:spectra}
\frac{dN}{dE} = f_0 \left(\frac {E}{E_0}\right)^{-\alpha},
\end{equation}
where $f_0$ is the normalization in units of
$\cm^{-2}\,\s^{-1}\,\TeV^{-1}$.
The gamma-ray energy flux is trivially related to $dN/dE$ by
$F_E=E dN/dE \propto E^{1-\alpha}$, from which we obtain a VHEGR flux,
\begin{equation}
F =  E_0 f_0 \int_{100\,\GeV}^{10\,\TeV} dE\,\left(\frac{E}{E_0}\right)^{1-\alpha}\,,
\end{equation}
and for sources with a measured redshift a corresponding
isotropic-equivalent luminosity, $L=4\pi D_L^2 F$, where $D_L$ is the
luminosity distance.
The resulting $f_0$, $E_0$, $\alpha$, $F$, and $L$ are collected in Table
\ref{tab:TeVsources}.  In addition we list the redshift, inferred
distance, and absorption-corrected
intrinsic spectral index at $E_0$, obtained via
\begin{equation}
\hat{\alpha}
=
-\left.\frac{d\ln E^{-\alpha} e^{\tau\left(E,z\right)}}{d\ln E}\right|_{E_0}
\simeq
\alpha - \tau\left(E_0,z\right)\,,
\end{equation}
where $\tau\left(E,z\right)$ is the optical depth accrued by a VHEGR
emitted at redshift $z$ and with observed energy $E$.\footnote{This
  differs subtly from the definition of $\tau_E(E,z)$ in Paper I,
  where there we set $E$ to the {\em emitted} energy of the gamma
  ray.  A full definition at arbitrary observer redshift can be found
  in Equation (\ref{eq:tau}).}
For high-redshift sources $\hat{\alpha}$ can be less than $2$,
implying that an intrinsic spectral upper-cutoff must exist.  Here we
conservatively take this to be at $E\simeq10\,\TeV$, which is well justified
given the distances to the two sources that dominate the observed local $\TeV$
flux, Mkn 421 and 1ES 1959+650, though as we shall see below, the heating rate
is relatively insensitive to the particular values of the lower and upper
spectral cutoffs.

\begin{deluxetable*}{lccccccccccl}\tabletypesize{\tiny}
\tablecaption{List of TeV Sources with Measured Spectral Properties in Decreasing $100\,\GeV$--$10\,\TeV$ Flux Order\label{tab:TeVsources}}
\tablehead{
\colhead{Name} &
\colhead{$z$} &
\colhead{$D_C$ \tablenotemark{a}} &
\colhead{$f_0$ \tablenotemark{b}} &
\colhead{$E_0$ \tablenotemark{c}} &
\colhead{$\alpha$ \tablenotemark{d}} &
\colhead{$F$ \tablenotemark{e}} &
\colhead{$\log_{10} L$ \tablenotemark{f}} &
\colhead{$\hat{\alpha}$ \tablenotemark{g}} &
\colhead{$\dot{q}$ \tablenotemark{h}} &
\colhead{Class \tablenotemark{i}} &
\colhead{\mbox{Reference\hspace{2.8cm}}}
}
\startdata
Mkn 421		 & 0.030   &  129    & 68    & 1        & 3.32  & $1.7\times10^{3}$ & 45.6 & 3.15       &  44     & H & \citet{Chandra+10} \\
1ES 1959+650	 & 0.047   &  201    & 78    & 1 	& 3.18 	& $1.6\times10^{3}$ & 45.9 & 2.90       &  47     & H & \citet{aharonian+03a} \\
1ES 2344+514	 & 0.044   &  190    & 120   & 0.5 	& 2.95  & $2.3\times10^{2}$ & 45.0 & 2.82       &   9.2   & H & \citet{albert+07c}\\
Mkn 501          & 0.034   &  150    & 8.7   & 1 	& 2.58  & 85                & 44.4 & 2.39       &   6.0   & H & \citet{Huang+09} \\
3C 279		 & 0.536   & 2000    & 520   & 0.2 	& 4.11 	& 68                & 46.9 & 2.53       &   1.0   & Q & \citet{magic+08} \\
PKS 2155-304	 & 0.116   &  490    & 1.81  & 1 	& 3.53 	& 64                & 45.4 & 2.75       &   1.4   & H & \citet{hess+10a} \\
PG 1553+113	 & $>0.09$ &  $>380$ & 46.8  & 0.3      & 4.46  & 41                & $>44.9$ & $<4.29$ & $<3.88$ & H & \citet{aharonian+08b}\\
W Comae		 & 0.102   &  430    & 20    & 0.4 	& 3.68	& 31                & 44.9 & 3.41       &   0.6   & I & \citet{acciari+09a} \\
3C 66A		 & 0.444   & 1700    & 40    & 0.3 	& 4.1	& 28                & 46.3 & 2.43       &   0.4   & I & \citet{acciari+09b}\\
1ES 1011+496	 & 0.212   &  870    & 200   & 0.2 	& 4 	& 26                & 45.5 & 3.66       &   0.4   & H & \citet{albert+07a} \\
1ES 1218+304 	 & 0.182   &  750    & 11.5  & 0.5 	& 3.07	& 24                & 45.4 & 2.37       &   0.8   & H & \citet{acciari+10b}\\
Mkn 180		 & 0.045   &  190    & 45    & 0.3 	& 3.25 	& 20                & 44.0 & 3.17       &   0.6   & H & \citet{albert+06a} \\
1H 1426+428	 & 0.129   &  540    & 2     & 1 	& 2.6 	& 20                & 45.0 & 1.71       &   1.4   & H & \citet{aharonian+02a} \\ 
RGB J0710+591 	 & 0.125   &  520    & 1.36  &  1 	& 2.69 	& 15                & 44.8 & 1.83       &   0.9   & H & \citet{acciari+10a} \\
1ES 0806+524	 & 0.138   &  580    & 6.8   & 0.4	& 3.6 	& 10                & 44.7 & 3.21       &   0.2   & H & \citet{acciari+09c} \\
RGB J0152+017 	 & 0.080   &  340    & 0.57  & 1 	& 2.95 	& 8.5               & 44.1 & 2.45       &   0.3   & H & \citet{aharonian+08a} \\
1ES 1101-232 	 & 0.186   &  770    & 0.56  & 1 	& 2.94 	& 8.2               & 44.9 & 1.50       &   0.3   & H & \citet{aharonian+07c} \\
1ES 0347-121 	 & 0.185   &  770    & 0.45  & 1 	& 3.1	& 8.2               & 44.9 & 1.67       &   0.3   & H & \citet{aharonian+07b} \\
IC 310           & 0.019   &   83    & 1.1   & 1        & 2.0   & 8.1               & 42.8 & 1.90       &   0.1   & H & \citet{alek_etal:10} \\
PKS 2005-489	 & 0.071   &  300    & 0.1   & 1 	& 4.0 	& 8.0               & 44.0 & 3.56       &   0.1   & H & \citet{aharonian+05a} \\
MAGIC J0223+430	 & --      &   --    & 17.4  & 0.3 	& 3.1 	& 7.6               & --   & $<3.1$     &   0.2   & R & \citet{aliu+09} \\
1ES 0229+200 	 & 0.140   &  590    & 0.7   & 1 	& 2.5 	& 6.4               & 44.5 & 1.51       &   0.5   & H & \citet{aharonian+07a} \\
PKS 1424+240	 & $<0.66$ & $<2400$ & 51    & 0.2	& 3.8 	& 6.3               & $<46.1$ & $>1.42$ &   0.1   & I & \citet{acciari+10c}\\
M87      	 & 0.0044  &   19    & 0.74  & 1 	& 2.31	& 5.9               & 41.4 & 2.29       &   0.6   & R & \citet{acciari+08a}\\
BL Lacertae	 & 0.069   &  290    & 0.3   & 1 	& 3.09 	& 5.4               & 43.8 & 2.67       &   0.2   & L & \citet{albert+07b} \\
H 2356-309	 & 0.165   &  690    & 0.3   & 1 	& 3.09 	& 5.4               & 44.6 & 1.86       &   0.2   & H & \citet{aharonian+06a} \\
PKS 0548-322     & 0.069   &  290    & 0.3   & 1 	& 2.86	& 4.0               & 43.7 & 2.44       &   0.2   & H & \citet{aharonian+10a} \\
Centaurus A	 & 0.0028  &   12    & 0.245 & 1	& 2.73	& 2.8               & 40.7 & 2.72       &   0.2   & R & \citet{raue+10}\\
\enddata
\tablenotetext{a}{Comoving distance in units of $\Mpc$}
\tablenotetext{b}{Normalization of the observed photon spectrum that we assume to be of the form $dN/dE=f_0 (E/E_0)^{-\alpha}$, in units of $10^{-12}\,\cm^{-2}\s^{-1}\TeV^{-1}$}
\tablenotetext{c}{Energy at which we normalize the spectrum, in units of $\TeV$}
\tablenotetext{d}{Observed spectral index at $E_0$} 
\tablenotetext{e}{Integrated energy flux between $100\,\GeV$ and $10\,\TeV$, in units of $10^{-12}\,\erg\,\cm^{-2}\s^{-1}$}
\tablenotetext{f}{Inferred isotropic integrated luminosity between $100\,\GeV$ and $10\,\TeV$, in units of $\erg\,\s^{-1}$}
\tablenotetext{g}{Inferred intrinsic spectral index at $E_0$}
\tablenotetext{h}{Local plasma-instability heating rate in units of $10^{-10}\eV\,\cm^{-3}\Gyr^{-1}$}
\tablenotetext{i}{H, I, L, Q, and R correspond to high-energy,
  intermediate-energy, low-energy peaked BL Lacs, flat spectrum radio quasars, and radio galaxies of Faranoff-Riley Type I (FR I), respectively.}
\end{deluxetable*}

\begin{figure}
\begin{center}
\includegraphics[width=0.95\columnwidth]{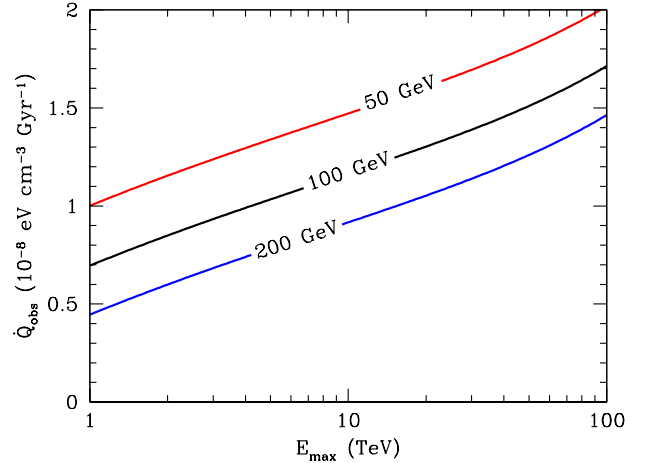}
\end{center}
\caption{Local heating rate due to only the observed TeV blazars as a
  function of upper-cutoff ($E_{\rm max}$) and lower-cutoff energies
  (red, black, and blue lines).  In particular, note that the overall
  value of $\dq_{\rm obs}$ is relatively insensitive to reasonable
  variations in the spectral cutoffs.}\label{fig:Qobs}
\end{figure}

Using the observed VHEGR source in Table \ref{tab:TeVsources}, the
heating rate associated with the 23 IBL/HBL blazars (labeled I and H in
the table, see below) is 
\begin{equation}
\dq_{\rm obs}
\simeq
\sum_{\rm obs~AGN} 
\frac{E_0^2 f_0}{\Dpp(E_0,0)} \int_{100\,\GeV}^{10\,\TeV} \frac{dE}{E_0} \left(\frac{E}{E_0}\right)^{2-\alpha}\,.
\end{equation}
Note that for $\alpha\simeq3$, characteristic of the two dominant $\TeV$
sources, this depends logarithmically upon the lower and upper
spectral cutoffs.  The resulting $\dq_{\rm obs}$ is shown in
Figure \ref{fig:Qobs} as a function of the upper spectral cutoff,
$E_{\rm max}$ for a handful of different lower cutoffs, ranging from
$50\,\GeV$ to $200\,\GeV$.  For our fiducial values,
$\dq_{\rm obs}=1.2\times10^{-8}\,\eV\,\cm^{-3}\,\Gyr^{-1}$,
with this only weakly depending upon the various cutoffs, changing by
at most by a factor of 2 over the reasonable ranges.  
Hence, for the remainder of this work, we choose a lower spectral cutoff of 100 GeV.

To complete this estimate, we now correct for the selection effects of
VHEGR observations.  To correct for the pointed nature of VHEGR
observations, we rely on the all-sky $\GeV$ gamma-ray observations
from the Fermi satellite \citep{Fermi_AGNCatalogue2010} of TeV
blazar. Those belong to 2 subclasses of blazars, namely high-energy
peaked BL Lacs (HBL) and the somewhat less efficient accelerators,
intermediate-energy peaked BL Lacs (IBL) which are, in some cases,
also able to reach energies beyond 100 GeV.  Outside of the Galactic
plane, \Fermi observes 118 high-synchrotron peaked (HSP) blazars and a
total of 46 high-synchrotron peaked (ISP) blazars.\footnote{The source
  classes of HSP/ISP are very similar to the commonly used HBL/IBL
  classes. Hence we identify both for the remainder of this work.}
Roughly half of the latter are likely to emit VHEGRs as indicated by
their flat spectral index between 0.1 and 100 GeV, $\Gamma\lesssim2$
(see the spectral index distribution of Figure 14 in
\citet{Fermi_AGNCatalogue2009}). Of these potential 141 TeV blazars,
only 22 have also been coincidentally identified as TeV
blazars (out of a total of 28 coincident TeV sources), while there are
a total of 33 known TeV blazars (29 HBL, 4 IBL).  If these 141 sources
are all $\TeV$ emitters, but have not been detected due to incomplete
sky coverage of current TeV instruments, then the selection factor is
$\eta_{\rm sel} = 141/33=4.3$.  In addition, the duty cycle of
coincident $\GeV$ and $\TeV$ emission is $\eta_{\rm duty}=33/22=1.5$.
Here we assume that the luminosity distribution of observed VHEGR
sources reflects the true distribution after correcting for the
effects of flux limitations in the observations.  That this may be
done is justified empirically in Section 5.1.2 of Paper I, and
demonstrated explicitly in Figure 5 of Paper I.  Thus, we may use
constant correction factors (independent of luminosity) to estimate
the true distribution.  Finally, by excluding the Galactic plane for
galactic latitudes $b<10\degr$, this is an underestimate by roughly
$\eta_{\rm sky}=1.17$.  Taken together, the true heating rate in our
``standard'' model is then:
\begin{multline}
\left.\dq\right|_{z=0}
\simeq 7\times10^{-8}
\left(\frac{\eta_{\rm sys}}{0.8}\right)\\
\times
\left(\frac{\eta_{\rm sel}}{4.3}\right)
\left(\frac{\eta_{\rm duty}}{1.5}\right)
\left(\frac{\eta_{\rm sky}}{1.2}\right)
\,\eV\,\cm^{-3}\Gyr^{-1}\,,
\label{eq:LocalHeatingRate}
\end{multline}
where $\eta_{\rm sys}$ is a remaining coefficient of order unity correcting
  for systematic uncertainties. These include corrections to the spectral
  model we currently use and the corrections to the completeness of the Fermi sample of HBLs and IBLs in accounting for the complete population of VHEGR sources. The fact that there are already 4 radio galaxies of
  Faranoff-Riley Type I (FR I), 2 flat spectrum radio quasars, and 2 starburst
  galaxies emitting VHEGRs implies that we are probably too
  conservative in accounting for all the VHEGR sources.  To explore the effects of the variation in the heating rate, we also adopt an ``optimistic'' heating model that is normalized to fit the observed IGM inverted temperature-density relation \citep{Viel+09}. This "optimistic" model has 
  $\eta_{\rm sys}=1.6$ yielding a heating rate of $\left.\dq\right|_{z=0} =
  1.4\times10^{-7}\,\eV\,\cm^{-3}\Gyr^{-1}$.

In estimating $\dq$ from Equation (\ref{eq:qdotest2}) we have made
a number of implicit assumptions.  First, we have assumed that the
observed distribution of TeV blazars locally is representative of the
average distribution at {\it any given point} in the Universe.  Since the
blazars that have been observed locally go out to $z\approx 0.5$, we
believe this is a relatively safe assumption.  Second, we have assumed
that the current sample of blazars is sufficiently flux-complete to
dominate the heating rate at Earth.  That is, we have assumed that the
effective flux limit of the current generation of Cerenkov telescopes
is low enough that it captures the bulk of the sources responsible for
the local heating of the IGM.  That this is the case is less clear,
and will be the subject of Section \ref{sec:homogeneity} and the
Appendix in some detail.  Unfortunately, since the constant of
proportionality relating the local TeV blazar and quasar luminosity
densities is obtained using the observed set of blazars, our two
estimates are strongly correlated.

Nevertheless, this provides a convenient lower limit upon $\dq$,
and we use this heating rate, which we denote the "standard" heating
rate, as the mean heating rate of any fluid element in the present-day
Universe.  Over a Hubble time, the total heat deposited per unit
volume is then roughly  
\begin{equation}
\left. u_{\rm tot}\right|_{z=0}
\simeq
\frac{\left.\dq\right|_{z=0}}{H_0}
\approx 
10^{-6}\,\eV\,\cm^{-3}\,,
\end{equation}
which is sufficient to raise the temperature of $1+\delta=0.1$ regions to
approximately $1.7\times10^5\,\K$, and thus dramatically alter the
thermal history of voids.

\subsubsection{Estimating the $z>0$ Heating Rate}\label{sec:z>0 heating}

Extrapolating the above estimate to $z>0$ requires understanding how
the TeV blazar properties and population have evolved, as well as the
evolution in $f(F_E,E,z)$.  Generally, the average heating rate is
given by
\begin{equation}
\dq = \int dV\,d\log_{10}L\,d\alpha\,d\Omega\,
\tilde{\phi}_B(z;L,\alpha,\Omega)
\frac{\Omega}{2\pi} \dot{q}\,,
\end{equation}
where $\tilde{\phi}_B(z;L,\alpha,\Omega,z)$ is the physical number
density of blazars at a given redshift per unit logarithmic
isotropic-equivalent luminosity, spectral index, and blazar jet
opening angle (we've assumed all jets are symmetric).  We make the
simplifying assumption that $\tilde{\phi}_B$ is separable into
components describing the evolving luminosity density distribution,
$\tilde{\phi}_B(z,L)$, and a static, unit-normalized spectral
distribution, $\varphi_B(\alpha,\Omega)$, where
$\tilde{\phi}_B(z;L,\alpha,\Omega)=\tilde{\phi}_B(z,L)\varphi_B(\alpha,\Omega)$.
This produces,
\begin{multline}
\dq = \int d\log_{10} L\, L \tilde{\phi}_B(z,L)\\
\times \int D^2 dD\,d\alpha\,d\Omega\,
\int dE\,
\frac{\theta}{\Dpp} f \frac{L_E e^{-D/\Dpp}}{L D^2}\,,
\end{multline}
where the bulk of the redshift and luminosity dependence is now
contained in the outer-most integral over $\log_{10} L$, and we have used
$\tau_E\simeq D/\Dpp$ (since the heating is dominated by nearby
objects, for simplicity we do not distinguish between different
cosmological distance definitions, though see the Appendix for a more
careful treatment).  If there were no redshift or flux dependence
in the remaining terms, it would be possible to simply normalize the
heating rate by $\left.\dq\right|_{z=0}$ and the estimated
TeV-blazar luminosity density, which we define to be
\begin{equation}
\tilde{\Lambda}_B(z)\equiv\int_{\log_{10}L_{\rm min}}^\infty d\log_{10}L\, L \tilde{\phi}_B(z,L)\,,
\end{equation}
where $L_{\rm min}\simeq3\times10^{42}\,\erg\,\s^{-1}$ is chosen so
that the plasma instabilities operate efficiently at all redshifts of
interest.  However, this is not entirely the case since $f(F_E,E,z)$
retains some dependence upon $z$, and thus we set
\begin{equation}
\dq = \frac{\tilde{\Lambda}_B(z)}{\tilde{\Lambda}_B(0)} \left.\dq\right|_{z=0} \Qcorr(z)\,,
\label{eq:GlobalHeatingRate}
\end{equation}
where $\Qcorr$ provides a correction due to the changing
strength of the pair beam dissipation mechanism in comparison to
inverse-Compton cooling.

\begin{figure}
\begin{center}
\includegraphics[width=0.95\columnwidth]{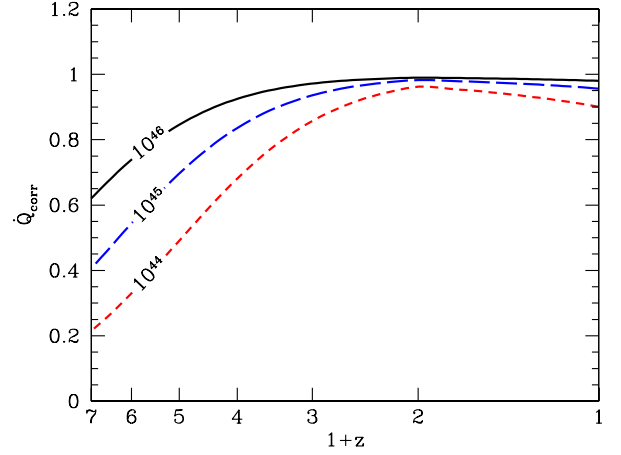}
\end{center}
\caption{Heating correction due to the intrinsic evolution of the
  plasma instability and inverse-Compton cooling rates as a function
  of redshift.  Shown are the corrections for
  $L=10^{46}\,\erg\,\s^{-1}$ (black solid),
  $10^{45}\,\erg\,\s^{-1}$ (blue long-dashed), and
  $10^{44}\,\erg\,\s^{-1}$ (red short-dashed).  For comparison, the
  TeV blazars that dominate the heating locally have luminosities of
  roughly $5\times10^{45}\,\erg\,s^{-1}$.}\label{fig:Qcorr}
\end{figure}

The VHEGR flux at Earth is dominated by a handful of very bright sources for which the absorption corrected $\alpha\simeq3$ at $1\,\TeV$ and
isotropic-equivalent luminosity $5\times10^{45}\,\erg\,\s^{-1}$.  This
simplifies the estimation of $\Qcorr$ significantly, giving
\begin{multline}
\Qcorr(z)
\simeq 
\iint dE dD  \frac{\theta E^{-2}}{\Dpp} e^{-D/\Dpp}
f\left( \frac{L_E e^{-D/\Dpp}}{4\pi D^2}, E, z \right)\\
\bigg/
\int dE \theta E^{-2}\,,
\end{multline}
in which the form and magnitude of $L_E$ is fixed.
The resulting correction factors are shown in Figure \ref{fig:Qcorr}
for isotropic-equivalent luminosities above $100\,\GeV$ ranging from
$10^{44}\,\erg\,\s^{-1}$ to $10^{46}\,\erg\,\s^{-1}$.
Generally, $\Qcorr$ makes a small ($\lesssim15\%$) correction following
the peak of the quasar luminosity function ($z=2$), though can grow to
as much as $80\%$ by $z=6$ for dim objects.  However, since we will be
primarily interested in bright blazars at $z\lesssim 4$, $\Qcorr$
modifies the heating rate at relevant redshifts by $\lesssim50\%$ (see also Section~\ref{sec:TBprop}).
Thus, given the comparatively larger uncertainties associated with the
TeV blazar luminosity function, plasma heating mechanism, and
additional potential VHEGR sources, we neglect $\Qcorr$ in what follows.

\begin{figure}
\begin{center}
\includegraphics[width=0.95\columnwidth]{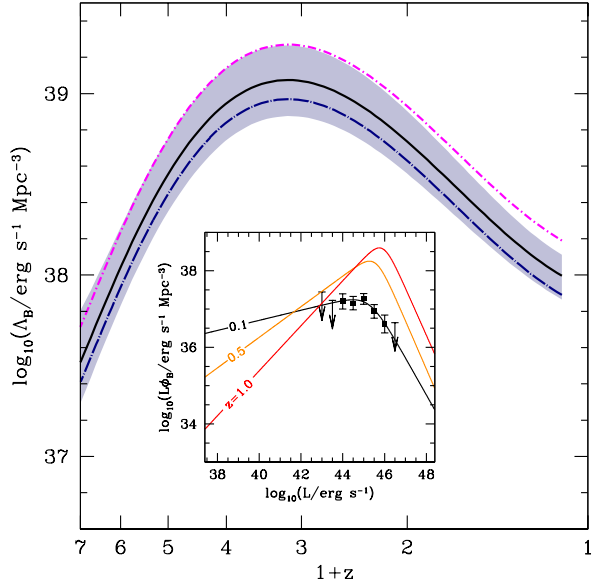}
\end{center}
  \caption{
    Blazar luminosity density (in comoving units, with $\eta_{\rm sys}=1$) as a
    function of redshift. The shaded region represents the 1-$\sigma$
    uncertainty that results from a combination of the uncertainty in
    the number of bright blazars that contribute to the local heating
    and in the uncertainties in the \citet{Hopkins+07} quasar luminosity density to which we
    normalize.
    Our optimistic and standard models, defined in
    Section \ref{sec:z=0 heating},
    are shown by the
    magenta short-dash-dot and blue long-dash-dot lines.
    Insert: Comparison between the observed blazar
    luminosity function (data points with one sigma error bars and
    upper limits) with the quasar luminosity function that has been
    shifted by a factor of $2.1\times 10^{-3}$ in luminosity density
    and $0.55$ in luminosity. This insert is a simplified reproduction
    of Figure 5 of Paper I. 
    \label{f:BLD}}
\end{figure}

Therefore, obtaining the heating rate as a function of $z$ is reduced
to estimating $\tilde{\Lambda}_B(z)/\tilde{\Lambda}_B(0)$, the normalized evolution of
the blazar luminosity density.
However, no
VHEGR emitting blazars are known beyond $z\simeq0.7$,
presumably due to the pair-creation absorption associated with
propagation through the EBL.  As a consequence, nothing is known about
$\tilde{\Lambda}_B(z)$ at high redshifts directly.  
Nevertheless, in Paper I we showed that a close relationship between
$\tBLF$ and the quasar luminosity function of \citet{Hopkins+07},
$\tQLF$, exists at low $z$: 
\begin{equation}
\tBLF(0.1,L) \simeq 3.8\times10^{-3}\,\tQLF(0.1,1.8L)\,.
\end{equation}
The insert in Figure \ref{f:BLD} shows a comparison between the
observed TeV-blazar luminosity function, calculated from the
distribution of the sources in Table \ref{tab:TeVsources} and applying
an empirically determined flux limit of
$4.19\times10^{-12}\,\erg\,\cm^{-2}\,\s^{-1}$, with $\BLF(0.1,L)$
(for a full description of how the luminosity function was formed, and
how it compares to $\QLF$, see Section 5 of Paper I).
Furthermore, we showed
that once the ICCs are suppressed by the plasma
beam instabilities (or some analogous mechanism), extending this
relationship to high-$z$ was in excellent agreement with the best 
current constraints upon the high-$z$ TeV blazar population: the \Fermi
TeV blazar statistics and the \Fermi measurement of the extragalactic
gamma-ray background (between $100\,\MeV$ and $100\,\GeV$).  Thus, we
estimate $\Lambda_B(z)\simeq2.1\times10^{-3}\Lambda_Q(z)$, shown in Figure
\ref{f:BLD}, where $\Lambda_Q(z)$ is the luminosity density of
quasars.

Figure \ref{f:BLD} shows that employing the quasar luminosity density to estimate the heating rates
has profound consequences.  In particular, the inferred luminosity density of
TeV blazars (solid line) rises rapidly with increasing $z$, with the comoving density
increasing by roughly a factor of $\sim 10$ by $z=2$ 
\citep[see also Figure 8 of ][]{Hopkins+07}.  In physical units this
corresponds to a increase by a factor of nearly $300$.
Thus, we expect an increase in the local heating rate by a similar
factor over the presently estimated value near $z=2$.\\

\subsection{Homogeneity of TeV-Blazar Heating} \label{sec:homogeneity}

We now investigate the assumption of even heating as a prelude to our
simple one-zone model of the IGM. 
In lieu of large-scale simulations, the homogeneity of the heating due
to TeV blazars is difficult to assess for a variety of reasons.
First, the duty cycle of the TeV blazars is unknown.  Second, the
density at high redshifts is poorly constrained.  Third, the
importance of clustering bias is unclear.  Fourth, it is difficult even to define
which blazars are relevant, e.g., which luminosity range contributes
the bulk of the local heating.
Nevertheless, we make an attempt to roughly characterize the possible
range in the stochasticity of the local heating rates via a number of
different estimates.

\subsubsection{Mean Separation of TeV Blazars}

\begin{figure}
\begin{center}
\includegraphics[width=0.95\columnwidth]{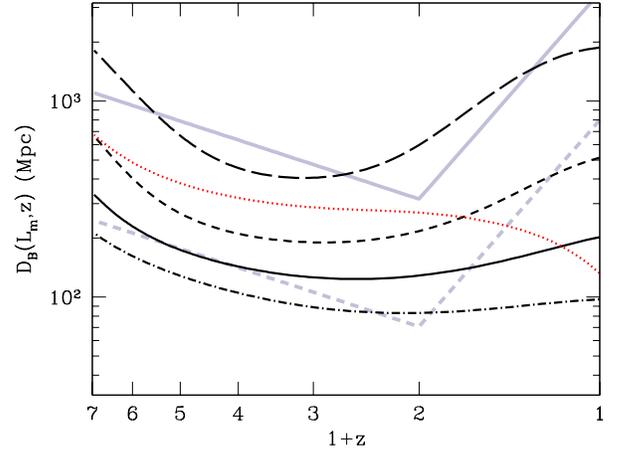}
\end{center}
\caption{Comoving mean separation of blazars with isotropic-equivalent
  luminosities ($L$, in the $100\,\GeV$--$10\,\TeV$ band) above
  $L_m=10^{43}\,\erg\,\s^{-1}$ (dot-dash), $10^{44}\,\erg\,\s^{-1}$
  (solid), $10^{45}\,\erg\,\s^{-1}$ (short-dash), $10^{46}\,\erg\,\s^{-1}$
  (long-dash), and the local luminosity-weighted median luminosity of
  TeV blazars (red dot), as functions of redshift.  For reference
  the spectrally averaged (with $\alpha=3$) mean free path, $\bDpp$, and the
  local mean free path for a $1\,\TeV$ gamma ray are shown by the
  grey solid and dashed lines, respectively.  For sources with a
  spectral break between $100\,\GeV$ and $1\,\TeV$, we anticipate the
  effective mean free path to lie between these.
}
\label{f:BMS}
\end{figure}

We begin by estimating the instantaneous comoving mean separation of
{\em visible} blazars
above various luminosity thresholds,
\begin{equation}
D_B(L_m,z) = (1+z)\left[\int_{\log_{10}L_m}^{\log_{10}L_M} \tBLF(L,z) \dlL\right]^{-1/3}\,,
\end{equation}
where the precise value of $L_M$ is unimportant as long as it is
above the peak of the luminosity function; here we choose
$L_M=2\times10^{46}\,\erg\,\s^{-1}$, consistent with the theoretically
expected upper limit of TeV blazar luminosities.  Since we have determined
$\tBLF$ from the observed blazar population, and are making use of the
isotropic-equivalent luminosities, this is independent of the blazar
jet opening angle; smaller opening angles will result in a
correspondingly larger number of objects such that the total number
seen is unchanged, and therefore $D_B$ is fixed.
A comparison between $D_B$ and the mean free path of VHEGRs is shown in Figure \ref{f:BMS}.
Because $\Dpp$ varies dramatically from $100\,\GeV$ to $10\,\TeV$, for
this purpose, we define a spectrally averaged mean free path, $\bDpp$,
determined implicitly by 
\begin{equation}
\int_{100\,\GeV}^{10\,\TeV} dE E^{1-\alpha} e^{-\bDpp(z)/\Dpp(E,z)}
=
e^{-1}
\int_{100\,\GeV}^{10\,\TeV} dE E^{1-\alpha}\,,
\end{equation}
in which we choose $\alpha=3$ based upon the local TeV blazar sample.
This is roughly the $e$-folding distance of the entire spectral band,
shown in Figure \ref{f:BMS} by the grey solid line,
and in practice is quite close to $\Dpp(225\,\GeV,z)$.  If the
intrinsic TeV blazar spectra peak above $100\,\GeV$, our
estimate of the typical VHEGR mean free path could be
significantly too large; thus we also compare $D_B$ to $\Dpp$ at
$1\,\TeV$ (the grey dashed line in Figure \ref{f:BMS}), providing an
extreme lower-bound upon the spectrally averaged mean free path in
practice.  
The rapid increase of the density of EBL photons with $z$, associated
with the larger star formation rate in the recent past, results in a
substantially reduced $\Dpp$ by $z=1$.  Prior to $z=1$, the physical
number density of EBL photons remains nearly constant, and thus
$\Dpp\propto(1+z)$ in comoving units.

In the present epoch, $\bDpp$ is quite large, and thus despite their
sparsity the mean separation of even bright blazars
($10^{46}\,\erg\,\s^{-1}$) is is less than $\bDpp$.
This remains true for objects with luminosities
$\le10^{45}\,\erg\,\s^{-1}$, which includes the local
luminosity-weighted median TeV blazar luminosity, $L_{0.5}(z)$,
defined such that
\begin{equation}
\int_{\log_{10}L_{0.5}(z)}^{\log_{10}L_M} L \tBLF(L,z) \dlL
=
0.5\tilde{\Lambda}_B(z).
\label{eq:L0.5}
\end{equation}
Hence, locally, we expect blazar
heating to be quite uniform.

At high redshift matters change somewhat.  Until $z=1$, $\bDpp(z)$ and
$\Dpp(1\,\TeV,z)$ both decrease more rapidly than $D_B$.  For $z>1$,
in comoving units $\bDpp(z)$ and $\Dpp(1\,\TeV,z)$ increase slowly,
though at a marginally larger rate than $D_B$.  As a result, near
$z\sim1$ the mean separation of TeV blazars is largest in comparison
to the VHEGR mean free path.  Thus, the mean separation between
objects more luminous than $10^{46}\,\erg\,\s^{-1}$ is larger than
$\bDpp$ near $z\sim1$.  Nonetheless, when the lower luminosity limit is
dropped to $10^{45}\,\erg\,\s^{-1}$ or below, generally
$D_B(z)<\bDpp(z)$ for all $z<6$.  Similarly, the mean separation
between blazars more luminous than $L_{0.5}(z)$ is also smaller than
our estimate of $\bDpp$ for the relevant redshift range.
Hence we may expect that nearly all patches of the Universe will be
illuminated by at least one luminous TeV blazar following $z\sim6$.

\subsubsection{Estimates of the Number of TeV Blazars that Contribute Significantly to the Heating Rate}

\begin{figure}
\begin{center}
\includegraphics[width=0.95\columnwidth]{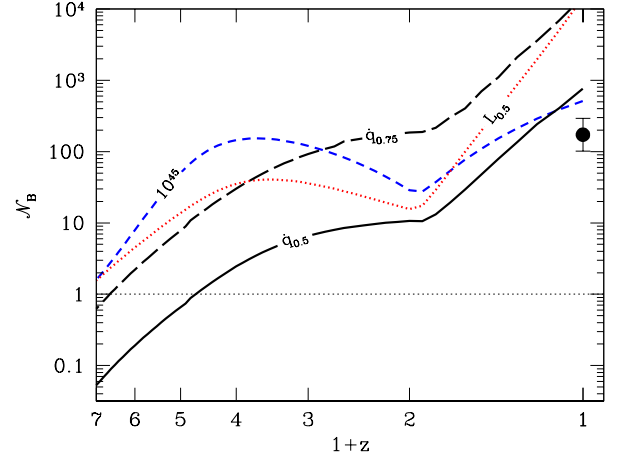}
\end{center}
\caption{Number of TeV blazars, estimated in various ways, that
  contribute to the heating of a given patch as a function of
  redshift.
  Definitions for ${\mathcal N}_B$ include:
  the number of blazars with intrinsic isotropic-equivalent
  luminosities above $10^{45}\,\erg\,\s^{-1}$ within the
  $\bar{\tau}=1$ surface (blue short-dash),
  the number of blazars with luminosities above that
  luminosity-weighted median value at each redshift within the
  $\bar{\tau}=1$ surface (red dot),
  and the number of blazars with individual heating rates that exceed
  that above which 50\% and 75\% of the total heating is produced
  (black solid and long-dash, respectively).
  For comparison, our estimate of the completeness-corrected number of
  TeV blazars that are presently observable {\em in the TeV} is shown
  by the filled back point, with error bars denoting the Poisson
  uncertainty only.
}
   \label{f:numBlazars}
\end{figure}

Closely related to the mean separation is the number of blazars within
the $\bar{\tau}\simeq D/\bDpp=1$ surface (defined explicitly in the 
Appendix) above some luminosity limit. 
Figure \ref{f:numBlazars} shows these for objects with luminosities
above $10^{45}\,\erg\,\s^{-1}$ (blue short-dash), and $L_{0.5}$ (red
dot).
While these may be roughly inferred from the associated mean
separations in Figure \ref{f:BMS}, they differ slightly from
$(4\pi/3)(\bDpp/D_B)^3$ due to the rapidly evolving $\bDpp$ and blazar
population combined with the finite look-back time in the integral.
At $z=0$ both are well above unity, 
with the luminosity-weighted median value substantially exceeding our
estimate of roughly 170 visible extragalactic TeV
blazars.\footnote{Recall that in Table \ref{tab:TeVsources} we have
  listed 23 TeV Blazars, but a correction factor of 7.5 needs to be
  applied to account for the incompleteness of present TeV
  surveys.} 
However, since generally $\bDpp(z)\simeq4.4\Dpp(1\,\TeV,z)$ not all of the
blazars that contribute to the local heating are expected to appear as
strong TeV sources.

The low-$z$ behavior of the number of blazars is dictated by the rapid
evolution in $\Dpp$, associated with the recent rapid variation in the
star formation rate (and thus the number density of EBL photons).
Prior to $z=1$, the EBL density, and hence $\Dpp$, is roughly constant
in physical units, and the evolution number of visible blazars becomes
indicative of the underlying evolution of blazar population.  At all
$z<6$ these estimates of the number of blazars responsible for the
bulk of the heating, ${\mathcal N}_B$, exceed unity, implying only a small
fractional spatial variation in the blazar heating rate.
However, while the luminosity-weighted median estimate of ${\mathcal N}_B$ does
give some idea of which population is responsible for most of the TeV
blazar luminosity density, neither encapsulates the population
responsible for the majority of the local heating.  Thus, if the local
heating were dominated by a handful of luminous sources, it is
possible for the stochasticity to be much larger.

Therefore, we show a third estimate of ${\mathcal N}_B$, corresponding to the
number of sources with $\dot{q}$s larger than the
heating-rate-weighted median value (see the Appendix for a precise definition).  
That is, at a given redshift, the 
${\mathcal N}_B$ sources with the highest $\dot{q}$s generate half of the total
heating rate.  Thus, if a single source were to dominate the local
heating rate, ${\mathcal N}_B$ would be considerably less than unity, indicating
a large degree of variability.  However, below $z\sim3.5$ this is not
the case; the fractional heating rate-defined ${\mathcal N}_B$ is considerably
larger than unity.  At high redshifts the heating rate becomes
increasingly dominated by more distant, luminous, and rarer objects,
departing from the previous estimate of ${\mathcal N}_B$.

Nevertheless, while for $z\gtrsim3.5$ few sources contribute nearly
half of the heating rate, the total heating rate must be comparable to
the total VHEGR luminosity density of the blazars.
Therefore, the total number of contributing objects must be similar to
the number of visible blazars above the luminosity-weighted median
$L$ (i.e., the red dotted line in Figure \ref{f:numBlazars}).  As a
consequence, the number 
of relevant sources must rapidly rise with decreasing fractional
heating rate.  This is explicitly indicated by the dashed line in
Figure \ref{f:numBlazars}, which shows the number of sources
responsible for 75\% of the local heating rate.

In any case, it is clear that for the redshifts at which blazar
heating is likely to be important, $z\lesssim4$, the heating rate will
be relatively uniform.  Between $z\sim4$ it may experience order 50\%
fluctuations, and by $z\sim6$ will exhibit order unity deviation.
However, we note that all of our estimates of the heating
inhomogeneity are predicated upon the assumption that the $\tBLF$
accurately represents the distribution of blazars, and has the
associated considerable uncertainties.
For a more thorough discussion of different estimates of the number of
contributing blazars, and a discussion of which blazars dominate the
heating rate at a given redshift, see the Appendix.

In addition to the statistics of blazars, the homogeneity also depends
upon the intrinsic variability of TeV blazars.  During the time that
the TeV blazars have been observed their VHEGR emission has remained
remarkably stable.  However, this provides only a weak lower limit
($\sim4\,\yr$) upon the variability timescale in these objects \citep{Derm_etal:10}.
In addition to depending upon the properties of the source itself, the
variability of TeV blazars depends upon the primary emission mechanism
responsible for the VHEGR component of TeV blazars.  There are two classes of
inverse-Compton models to explain the VHEGR emission:
\begin{enumerate}
\item {\em Synchrotron self-Compton (SSC) model:} In this model, the
  synchrotron radiation field is Compton up-scattered to TeV energies
  by a relativistic electron population
  \citep{Jones+1974, Ghisellini+1989}.  Recent work has led to the
  conclusion that a simple homogeneous, one-zone, SSC model cannot
  explain the SED of the majority of blazars \citep[see Figure 36 of][]{Fermi_SED2010}.
  However, models with multiple SSC components are consistent with the
  blazar SEDs.  Typically these invoke a steady, primary SSC component
  which peaks at the IR/optical (S) and $\gamma$-ray band (IC), and a
  second more energetic and usually more variable component, which
  peaks in the UV or X-ray band (S) and at GeV/TeV energies (IC).
  The variability of this energetic component increases the
  probability that a given patch of the IGM will see a TeV blazar
  during its history, resulting in larger homogeneity in the blazar
  heating.
\item {\em External radiation Compton (ERC) scenario:} this model
  proposes that the relativistic jet electrons Compton scatter an
  external radiation field \citep{Sikora+1994, Dermer+2002} from the
  accretion disk or the dusty torus surrounding it. In the first case,
  the disk generates UV seed photons which are then reflected toward
  the jet by the broad line region within a typical distance from the
  accretion disk of the order of 1 pc. In the second case, the dusty
  torus could provide IR seed photons that are emitted at larger
  distances from the jet.  In any of these ERC models, the VHEGR
  emission is expected to be very steady, implying 
  long-lived TeV blazars, resulting in more patchy blazar heating.
\end{enumerate}

We leave detailed studies of the inhomogeneity in the blazar heating rate
resulting from intrinsic inhomogeneity and variability in the spatial
distributions of the blazars themselves for future work.\\

\subsection{Properties of the Blazars Responsible for the Heating}
\label{sec:TBprop}

\begin{figure}
\begin{center}
\includegraphics[width=0.95\columnwidth]{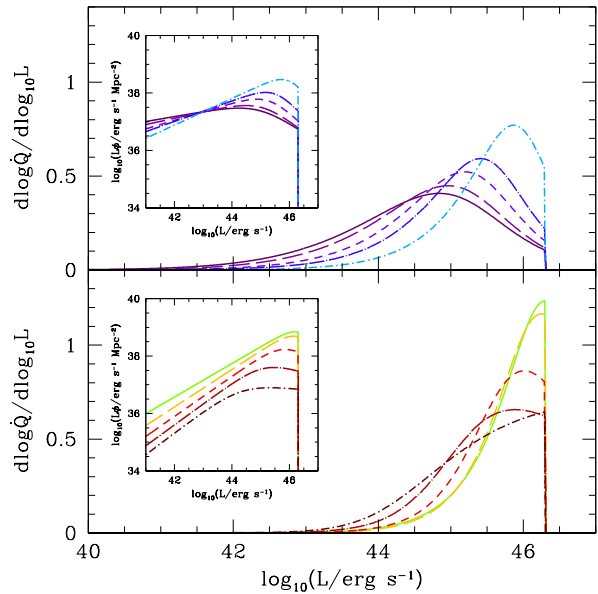}
\end{center}
\caption{$d\log\dq/d\log_{10}L$ as defined in Section
  \ref{sec:RTBs} for the TeV blazars responsible for half of the total
  heating rate for a number of redshifts (top: from violet blue to light-blue:
  $z_o=0$ (solid), $0.1$ (long-dash), $0.3$ (short-dash), $0.5$
  (long-dash-dot), and $1.0$ (short-dash-dot); bottom: from green to
  dark-red: $z_o=2$ (solid), $3$ (long-dash), $4$ (short-dash), $5$
  (long-dash-dot), and $6$ (short-dash-dot)).  For reference,
  $L\BLF(z_o,L)$ is shown in the inset (in comoving units).
}
\label{f:dQdL}
\end{figure}

In our discussion of the homogeneity of blazar heating we have
necessarily attempted to define a class of objects responsible for the
bulk of the heating. This is most directly done by considering those TeV
sources which produce, say, 50\% of the total heating rate at a given
observer redshift.  Given this population, we may also now address the
properties of the most relevant sources themselves; i.e., we can
identify which types of TeV blazars dominate the heating.  As we have
already mentioned, this cannot be too different from the set of
blazars which dominate the luminosity density.  We address this
question in some detail in the Appendix, including constructing a
simple analytical toy model.

Figure \ref{f:dQdL} shows the luminosity distribution of the
heating-rate integrand once integrated over distance and assuming our
form for $\tBLF(z,L)$ (see the Appendix for a precise definition).
The upper-half of the heating rate is dominated by sources with
luminosities of approximately $10^{45}\,\erg\,\s^{-1}$ at low
redshifts, and rises to $10^{46}\,\erg\,\s^{-1}$ by $z\sim2$ before
declining slightly thereafter.  This evolution is driven both by the
changing shape and the changing normalization of $\tBLF$ with
redshift.  At all redshifts this is larger than the median luminosity.
Thus, generally it appears that the heating is due predominantly to
high-luminosity objects.  Due to the relatively small $\bDpp$, these
are also necessarily nearby.

\subsection{Other Potential VHEGR Sources}\label{sec:alternatives}

Up to now, our focus on the TeV-blazars is motivated by the fact that the vast majority of extragalactic VHEGR sources observed are blazars.  Other potential source of VHEGR emission exist and so we will discuss these sources in this section.  In assessing their potential contribution in VHEGRs, however, we find that these VHEGR sources are subdominant.

\subsubsection{Starburst Galaxies}

Starburst galaxies are characterized by the presence of rapid star
formation, and hence, are sites of numerous supernovae.
Supernovae are known to be efficient particle accelerators, and are
presumed to be the primary source of the galactic cosmic rays.  
As cosmic rays can produce VHEGRs, starburst galaxies are potentially bright VHEGR sources.

While many starburst galaxies contain active AGN, that may also emit VHEGRs, unambiguous VHEGR emission form the starburst itself has been seen in at least two cases, M82 and NGC 253, with $E>100\,\GeV$ luminosities of
$5\times10^{39}\,\erg\,\s^{-1}$ and $2\times10^{39}\,\erg\,\s^{-1}$,
respectively \citep{Acci_etal:09,Acer_etal:09}.  The luminosity of these two sources are dwarfed by the local TeV blazars, and thus the VHEGR flux from 
M82 and NGC 253 does not contribute to the heating of the local IGM.  In addition, their VHEGR luminosity produces a pair beam which is too dilute for plasma processes such as the "oblique" instability to beat cooling via inverse Compton.  Hence, they do not contribute to the local VHEGR flux.

However, very high star-forming systems such as the ultra-luminous infrared
galaxies (ULIRGs) where $\LIR>10^{12}L_\odot$) and $\LIR$ is the bolometric infrared luminosity ($5\,\mum<\lambda<1000\,\mum$),
may evade this constraint.
If the VHEGR emission from starburst galaxies is due to cosmic rays
accelerated by supernovae, the
VHEGR luminosity above $100\,\GeV$, $L$ is then proportional
to the star formation rate.  For starbursts, this
is linearly related to the continuum infrared luminosity
\citep{Kenn:98}.  Thus, normalizing by M82 and NGC 253, we have
\begin{equation}
L\simeq6\times10^{40} \left(\frac{\LIR}{10^{12}L_\odot}\right)\,\erg\,\s^{-1}\,.
\end{equation}
While this relationship is extremely uncertain (the normalization
varies by a factor of two between M82 and NGC 253), it suggests that
ULIRGs, or perhaps hyper-luminous infrared galaxies (HLIRGs,
$\LIR>10^{13}L_\odot$) may be sufficiently bright to contribute to the
heating of the IGM.

While the fluxes from the brightest ULIRGs remain much smaller than those associated
with the typical TeV blazars, there are many more starburst galaxies than AGN.
At all redshifts ULIRGs constitute the high-luminosity tail of the
star-forming galaxy luminosity functions
\citep{LeFl_etal:05,Capu_etal:07,Magn_etal:09,Goto_etal:10}.
In the present epoch, the density of ULIRGs is roughly
$4\times10^{-7}\,\Mpc^{-3}$, and thus the corresponding VHEGR
luminosity density of these objects,
$\sim2\times10^{34}\,\erg\,\s^{-1}\,\Mpc^{-3}$
\citep[see Table 8 of ][]{Capu_etal:07}.
is negligible in
comparison to the TeV blazars, roughly $5\times10^{37}\,\erg\,\s^{-1}\,\Mpc^{-3}$.

However, due to the steep decline of the
luminosity function in the ULIRG range, small changes in the location
of the break luminosity result in large changes in the comoving
luminosity density of ULIRGs.  As a consequence, the comoving
luminosity density of ULIRGs grows much faster than that of quasars
(and thus presumably the TeV blazars).  Nevertheless, even at
$z=2$, roughly the redshift of peak star formation, the comoving
number density of ULIRGs remains below $\sim2\times10^{-4}\,\Mpc^{-3}$
\citep[see Table 8 of ][]{Capu_etal:07}, corresponding to a comoving
VHEGR luminosity density of
$\lesssim10^{37}\,\erg\,\s^{-1}\,\Mpc^{-3}$, more than two orders of
magnitude smaller than the contemporaneous TeV blazar population.  For
this reason, we neglect the starburst contribution to heating the IGM
here, though they may represent an
secondary source class.

\subsubsection{Magnetars \& X-ray Binaries}
Stellar-mass objects, such as magnetars and X-ray binaries, generally
have difficulty reaching the flux limits required for plasma cooling
to dominate the pair beam evolution.  Even at Eddington-limited VHEGR
luminosities, reaching isotropic-equivalent luminosities of
$10^{42}\,\erg\,\s^{-1}$ requires beaming factors of roughly $10^4$,
corresponding to jet opening angles of roughly $2^\circ$.  In
practice, the formation of radio jets is believed to be associated
with substantially sub-Eddington accretion flows in these objects, and
thus exacerbating the beaming requirement.  More importantly, these
sources may suffer compactness problems: the VHEGR emission regions in
stellar-mass jets must necessarily be very far from the central object
to avoid in situ pair-production.  Finally, were X-ray binaries and
magnetars generally strong, persistent VHEGR emitters with sufficiently
large fluxes, we would expect that many would have been already
detected.

\subsubsection{Gamma-ray Bursts}
Gamma-ray bursts (GRBs) are natural candidates due to their large
luminosities and strong inferred beaming, and we consider them as a
example of the general class of energetic transient sources.  
Unfortunately, little is known about the VHEGR emission of GRBs.
Presently there is a single report of a TeV signal associated with a
GRB \citep[GRB 970417A, ][]{Atki_etal:00,Atki_etal:03}, though due to
the large distances at which they can be observed 
and comparative rarity this is not unexpected.  However, such 
high-energy emission is possible in principle, presumably due to
inverse-Comptonization of the prompt emission and/or X-ray afterglow
\citep[see Section VIII of][and references therein]{Pira:04}. 
Fermi observations of GRBs have shown that for many events the
Band spectrum can be extended to $\sim100\,\GeV$ \citep{Fermi_GRB080916C:2009,Fermi_GRB090902B:2009,Fermi_GRB080825C:2009,Fermi_GRB090217A:2010}, though in at least
one case a spectral break below $10\,\GeV$ has been observed
\citep[GRB 090926A, ][]{Fermi_GRB090926A:2011}.  Thus it remains
unclear if in practice the high-energy emission is attenuated within
the emission region.
Moreover, since GRBs are inherently short-lived events, the luminosity
limits described in Paper I, which require the VHEGR
emitting phase to last for a plasma cooling timescale (roughly
$10^2\,\yr$--$10^3\,\yr$), are not directly applicable.  A similar analysis,
obtained by limiting the beam growth time to the GRB duration, gives
VHEGR isotropic-equivalent energies of $10^{54}\,\erg$.  This is
comparable to
the total prompt and afterglow emission for only the brightest bursts,
comprising roughly 5\% of GRBs observed by Swift \citep{Gehr-Rami-Fox:09}.  Nevertheless, even
assuming that all GRBs produce the requisite high energy emission, at
an optimistic present local rate of roughly
$0.5\,\Gpc^{-3}\,\yr^{-1}$, produces a comoving luminosity density of
$\lesssim10^{36}\,\erg\,\s^{-1}\,\Mpc^{-3}$, roughly three orders of
magnitude less than that due to TeV blazars at $z=1$.

\section{The Thermal History of the IGM}\label{sec:thermal history}

The previous sections have shown that the VHEGR
emission from luminous TeV blazars heats the IGM, quantified the
magnitude and stochasticity of this heating locally, and estimated its evolution as a
function of redshift.
We are now in a position to discuss its impact on the thermal history
of the IGM in detail.  In the following we will show that TeV blazar
heating can be substantial, dominating late time photoheating, and
that its uniform nature naturally imprints its signature onto the
temperature-density relation of the IGM.

The canonical history of the IGM is shaped by two
important events: H reionization by the first stars at
$z\sim6$--$10$ and HeII reionization by quasars at $z\sim3$
\citep[see ][ and references therein]{Furlanetto08}.  Hydrogen and HeII
reionization both heated the IGM to a
$T\sim 2\times10^4\,\K$--$3\times 10^4\,\K$ or higher (in the case of
HeII).  Subsequently, the Universe cooled via adiabatic expansion,
balanced by continuing photoheating due to ionization of recombining H.  The
entire canonical history of the IGM
can thus be summarized as a competition
between photoheating and adiabatic cooling, punctuated by intervals of
sudden photoheating.\footnote{We have ignored gravitational (shock) heating, which is progressively more important at densities larger than the mean density.}

The addition of TeV blazar heating adds an additional extended heating component, and fundamentally alters the canonical picture for the
thermal history of the IGM.  In the following we explore the
consequences of blazar heating using the one-zone model originally due to
\citet[hereafter HG97]{Hui97} \citep[see also ][]{Hui03}.  We begin
by introducing this model in detail (Section \ref{sec:model}),
describe thermodynamic consequences of the new heating contribution
from blazars and relate this to high-$z$ \Lya measurements 
(Section \ref{sec:IGM blazar heating}), and close with a discussion of
the implications for the local \Lya forest (Section \ref{sec:Lyalocal}).

\subsection{One-Zone Model for the IGM} \label{sec:model}
The thermal evolution of a fluid element in the IGM is governed by 
\begin{equation}\label{eq:T evolution}
 \frac {dT}{dt} = - 2 H T + \frac {2 T}{3(1+\delta)} \frac{d\delta}{dt} - \frac{T}{\Sigma \tilde{X}_i} \frac {d\Sigma \tilde{X}_i}{dt} + \frac 2 {3k_B \nbaryon} \frac {dQ}{dt},
\end{equation}
where $H$ is the redshift dependent Hubble function,
$\delta$ is the
mass overdensity, $\tilde{X}_i = n_i/\nbaryon$ is the {\it proper} number
fraction of species $i$, relative to the proper number density of
baryons, $\nbaryon=\Omega_B\rho_{\rm cr}/m_p$, $\rho_{\rm cr}$ is the critical mass density of the universe, and $dQ/dt$ is the heating and cooling rate of the
gas.  The heating and cooling of IGM gas is governed by four
processes: adiabatic cooling/heating from Hubble
expansion/gravitational collapse, H/He photoionization heating, H/He
recombination cooling, Compton cooling, and heating from TeV blazars.
The evolution of the proper number fraction of the various species is
given by 
\begin{equation}\label{eq:x evolution}
  \frac{d\tilde{X}_i}{dt} = -\tilde{X}_i\Gamma_i + \sum_{j.k}\nbaryon \tilde{X}_j\tilde{X}_k R_{ijk}\,,
\end{equation}
where the $\Gamma_i$ are the associated atomic rates, not to be
confused with the beam instability cooling rates discussed in Section
\ref{sec:pair propagation}.
Finally, we demand a prescription for the density evolution, i.e., the
evolution of $\delta$.  For this, we follow HG97 and assume the
Zel'dovich approximation: 
\begin{equation}\label{eq:delta evolution}
 1 + \delta = {\rm det}^{-1}\left(\delta_{ij} + D_+\psi_{ij}\right),
\end{equation}
where $D_+$ is the linear growth factor \citep{Peebles1980}.  The
$3\times 3$ matrix, $\psi$, is determined by initial conditions.  The
exact form of this matrix is irrelevant.  What is important, however,
is the probability distribution of the eigenvalues of this matrix.
For a Gaussian random field, the solution is known
\citep{Doroshkevich70}; we use the formulation of
\citet{Reisenegger95}.

Reionization of the Universe occurs in two stages:  first H is
reionized at some large redshift by stars and later He is
reionized around a redshift of $z \approx 3$ by quasars
\citep[see, e.g., ][]{Furlanetto08}. To model this
reionizing history, we adopt a sudden H photoionizing model (HG97): 
\begin{equation}\label{eq:sudden}
 J(z) = \left\{
\begin{array}{rl}
J_0 & {\rm for\ } z \leq z_{\rm reion}, \\
0   & {\rm for\ } z > z_{\rm reion},
\end{array}
\right.
\end{equation}
where $z_{\rm reion}$ is the redshift of H reionization and $J_0$ is
the normalization of the reionizing radiation.  While, this is not a
realistic model of how H reionization occurs, its late time evolution
(especially after $z\sim 3$) should be reasonably accurate.  This is
because the photoionizing background is observed to be roughly
constants at these redshifts \citep{Bolton05,Becker07,FG08} and the
late time temperature asymptotes to a single value.  This ``loss of
memory'' of the specific reionization history in the evolution of the
IGM is typical of reionization calculations \citep{Hui03}.  

To model the redshift dependence of H and He photoionization, we use the following
spectral model for the radiation: 
\begin{equation}\label{eq:spectral}
 J_{E}(z) = J(z)\left(\frac{E}{E_{\rm HI}}\right)^{-1.6}\left\{
\begin{array}{rl}
1 & {\rm for\ } E \leq E_{\rm He II}, \\
0.0  & {\rm for\ } E > E_{\rm He II}~\&~z > z_{\rm He}, \\
1.0  & {\rm for\ } E > E_{\rm He II}~\&~z < z_{\rm He}, \\
\end{array}
\right.
\end{equation}
where $E$ is the energy of the photon, $E_{\rm HI}$, $E_{\rm HeI}$,
$E_{\rm HeII}$ are the threshold energies corresponding to the
ionization of HI, HeI, and HeII, $z_{\rm He} = 3.5$ is the redshift of
He reionization. The spectral index of $-1.6$ is typical of quasars
and the spectral model of Equation (\ref{eq:spectral}) is similar to
the He reionization model studied by \citet{Furlanetto08}.  Our model
differs from theirs in that we ignore the density dependent effects of
He reionization, i.e., that dense regions are reionized first.
However, since we are interested solely in the magnitude of the effect
of blazar heating, adopting this simplified model is justified.

The normalization of the photoionizing background in our model is
fixed, determined by the H photoionizing rate: 
\begin{equation}
\Gamma_{\rm HI} = 4\pi \int_{E_{\rm HI}}^{\infty} J_E\sigma_{\rm HI} \frac {dE}{E},
\end{equation}
where $\sigma_{\rm HI}$ is the photoionizing cross section of HI.  We
choose the normalization of $J_E$ to be $\Gamma_{\rm HI}=5\times10^{-13}$, which is inferred (with significant uncertainty) from the mean absorption of the \Lya\ forest \citep[see for instance][]{Bolton05,FG08}. 
\begin{figure}
  \begin{center}
    \includegraphics[width=\columnwidth]{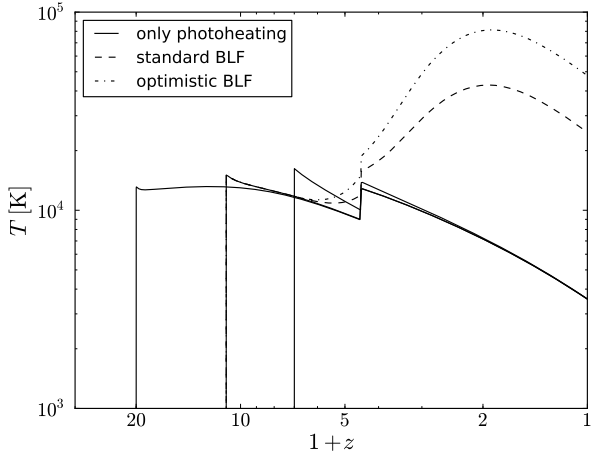}
  \end{center}
  \caption{Thermal history of the a $\delta = 0$ patch of the IGM for the
    numerical solution (solid lines) of Equations (\ref{eq:T evolution}),
    (\ref{eq:x evolution}), and (\ref{eq:delta evolution}) using the
    prescription for the microphysics as specified in the Appendix of HG97.  The
    solid curves are for sudden reionization histories for H and He (Equations
    (\ref{eq:sudden}) and (\ref{eq:spectral})) for $z_{\rm reion} = 19, 10, $
    and $6$ and $z_{\rm HeII} = 3.5$ going from left to right.  The dashed
    (dashed-dotted) lines show the evolution using the blazar luminosity density, 
    i.e., using the quasar luminosity density
    from \citep{Hopkins+07} to normalize the local heating rate in our standard
    (optimistic) models (see Figure~\ref{f:qlf EOS} and surrounding
    discussion).}
   \label{f:thermal history} 
\end{figure}

Equations (\ref{eq:T evolution}--\ref{eq:spectral})
constitute a complete model for the evolution of a fluid element in
the IGM.  We numerically integrate these equations using a
prescription for the heating and cooling microphysics specified in the
Appendix of HG97.  In Figure \ref{f:thermal history}, we plot the
evolution of the temperature for $\delta = 0$
patch with 
$z_{\rm reion} = 19,\, 10,$ and $6$.  The solid lines are the purely
photoionized models without the effect of additional heating. For each
of these models, we also set the redshift of HeII reionization at
$z_{\rm He II} = 3.5$, which results in a temperature jump at that
redshift.  We note that at late redshift, the three different (purely
photoionized) reionization histories asymptote to a single temperature
evolution, highlighting the "loss of memory" property that is generic to photoionization-dominated
models \citep{Hui03}.

\begin{figure}
  \begin{center}
    \includegraphics[width=\columnwidth]{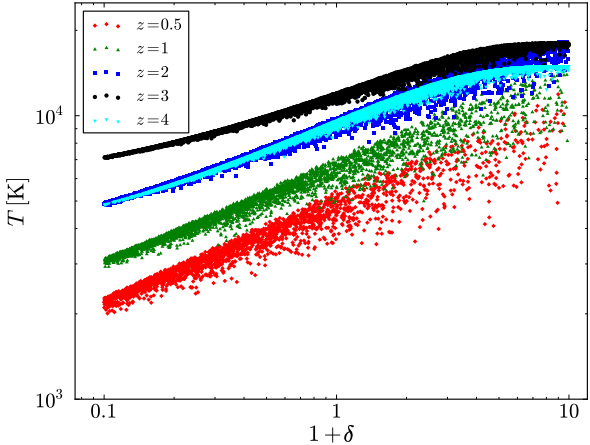}
  \end{center}
  \caption{Temperature-density scatter plot for $\approx 4000$
    realizations at $z=3$ (black dots), $z=2$ (blue squares), $z=1$ (green
    triangles), and $z=0.5$ (red diamonds) when heating from TeV
    blazars is ignored. The shape of the temperature-density plot can
    be clearly be fit with a power law (with a positive index) and
    steadily falls from $z=3.5$, i.e., after He reionization. 
    The correspondence between the $z=4$ and $z=2$ points are simply
    an accident of choosing He II reionization at $z=3.5$ and plotting
    the temperatures and densities at $z = 2$ and $4$.  Namely the Universe
    adiabatically cools, but the injection of heat at $z=3.5$ resets the
    temperature and it proceeds cooling from that point onward.} 
   \label{f:EOS}
\end{figure}

Generally, the thermal history of a given patch depends upon the
  particulars of the $\delta$-evolution of the patch.
  In Figure \ref{f:EOS} we show the temperature of $\approx4000$
  realizations of an evolving patch as a scatter plot at a number of
  redshifts, ranging from $z=0.5$ to $z=4$.
From this
it is clear that the temperature-density relation is well approximated
by a power law, consistent with those of HG97. Figure \ref{f:EOS}
represents a typical temperature-density relation, i.e., $T-\delta$
relationship, that is typical of most reionization calculations.  Low
density regions in the Universe are cooler compared to high-density
regions due to decreased recombination (and hence photoheating) and a
more rapid expansion (and hence greater adiabatic cooling).  Missing
from this simple picture are the effects of shocks and outflows from
galaxies, i.e., feedback (see \citealt{Dave_etal:10} for instance).
In addition, the temperature of the IGM is relatively cool ($\lesssim
10^4$ K) for $\delta \lesssim 0$.  This is because the temperature is
suddenly raised to a few $\times 10^4$ K after H and He reionization,
but rapidly cools due to the effects of adiabatic expansion.  These
generic characteristics are typical of most reionization models and are
expected following the arguments of HG97 and \citet{Hui03}.

\subsection{Contribution of TeV Blazar Heating}\label{sec:IGM blazar heating}

When the effects of heating due to the VHEGR emission from blazars are
included, the properties of the IGM are substantially altered.
The consequence of
TeV blazar emission is shown for a $\delta=0$ patch of the IGM by the dashed
line in Figure \ref{f:thermal history}.  The thermal history begins to deviate
significantly from that due to photonionization and adiabatic expansion alone by
$z\simeq6$, becomes dominant near $z\simeq3$, and peaks at roughly
$4$--$8)\times10^4\,\K$ at $z\simeq1$ before the rapid decline in $\Lambda_B(z)$
combined with adiabatic cooling causes the temperature to fall off (the range
corresponds to the uncertainty in estimating the number of blazars contributing
to the heating rate).  Thus, it is clear that heating by blazars is significant, dominates
at low redshifts (following He reionization), and potentially dominates the
thermal evolution of the IGM in low-density regions.

The effect of TeV blazar heating qualitatively changes the picture of the IGM.  
First, the temperature-density relation is inverted with the low-density regions being the hottest.  
Second, the overall temperature of the IGM is significantly hotter.
The reasons for both of these are twofold.  First, TeV blazars 
are a substantial reservoir of heating, potentially increasing the IGM
temperature by a few $\times 10^4\,\K$, and dominating the
contribution from ionizing photons for $1+\delta\lesssim10$.  Second,
the heating rate is nearly independent of density, depending most
strongly upon the number density of TeV blazars and the number density
of UV photons, both of which are nearly uniform 
(though see Section \ref{sec:homogeneity}).
The effect of a uniform heating rate is that the energy 
deposited per baryon is substantially larger in more tenuous regions
of the Universe, with underdense regions experiencing larger
temperature increases as a result.

\begin{figure}
\begin{center}
\includegraphics[width=\columnwidth]{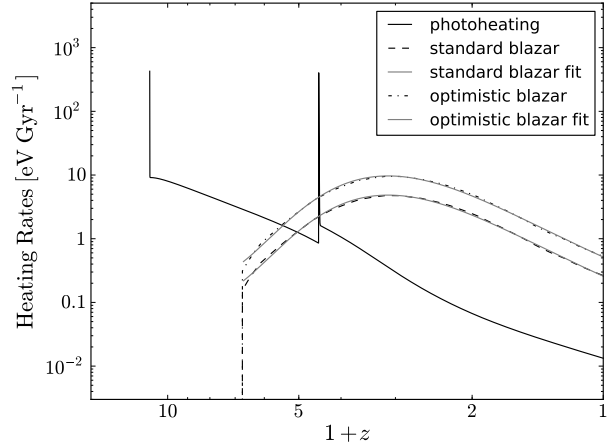}
\end{center}
  \caption{Photoheating (black solid line) and blazar heating rates for the standard (dashed line) and optimistic (dot-dashed line) as a
    function of redshift.  The fits (eq.(\ref{eq:fit})) are denoted by
    gray lines for the standard and optimistic blazar heating
    models. We use a $\delta = 0$ fluid element in the
    IGM and calculate the amount of heat per baryon (in units of eV)
    per Gyr.  The sharp jump at $z=3.5$ is due to HeII reionization.
    Aside from this point, it is clear that blazar heating dominates
    the heating of the IGM. 
    At late times $z\lesssim 2$, it is larger by over a factor of $10$, though this is substantially greater for underdense regions.}
   \label{f:rates}
\end{figure}

The dominance of TeV blazar heating over photoheating is shown explicitly in Figure \ref{f:rates}, where the blazar
heating rate (dashed line) is compared to the photoheating rate (solid
lines) as a function of redshift for a $\delta=0$ patch of the IGM.  
At $z=3.5$, there is a sudden jump in the photoheating rate due to
nearly instantaneous HeII reionization.  Following HeII 
reionization, the blazar heating rate is about an order of magnitude larger
than that due to photoheating\footnote{Prior to HeII reionization, the
  photoionization and blazar heating rates are inconsistent due to the
  artificial ionizing photon distribution assumed in Equation
  \ref{eq:spectral}.  Specifically, for reasons of simplicity, we have ignored the
  ionizing photons produced by the quasars.  Following HeII
  reionization, however, this is no longer an issue.}.  Because the
photoheating rate is $\propto(1+\delta)$, the dominance of TeV blazar
heating is even more apparent at lower densities.

The contribution of TeV blazar heating to the thermodynamics of the IGM for the standard model appears to be significant around the period of HeII reionization.  In our model, this is partially a result of our sudden reionization prescription for HeII reionization at $z=3.5$.  However, we can also show with the following order-of-magnitude calculation that the effect of TeV blazar heating must begin to be important around the era of He II reionization that has been observationally constrained to be around $z\sim 3$.  

To begin we first show that He reionization finishes around $z\approx 3$.  The {\it comoving} number density of He is 
\begin{equation}\label{eq:nHe}
n_{\rm He} = f_{\rm He}\frac {\Omega_B \rho_\rmn{cr}}{A_{\rm He}m_p} \approx 1.5\times10^{-8}\,{\rm cm}^{-3},
\end{equation}
where  
$f_{\rm He} = 0.24$ and $A_{\rm He} = 4$ are the primordial mass fraction and the atomic number of He, respectively.  To estimate the comoving density of HeII ionizing photons at $z\sim 3$, we note that the $1\ \Ry$ photon comoving density at $z=3.5$ from quasars in the \citet{Hopkins+07} QLF is $\dot{n}_{1\,\Ry} \approx 5\times 10^{-7}{\rm cm^{-3}\,Gyr^{-1}  }$ (e.g. see Figure 9 of \citealt{Hopkins+07}).  As the spectral index of quasars is $-1.6$, this implies that comoving number density of ionizing HeII photons is $\dot{n}_{4\,\Ry} \approx 5\times 10^{-8}{\rm cm^{-3}\,Gyr^{-1}}$. Thus, the total comoving density of HeII ionizing photons produced at $z=3.5$ is 
\begin{equation}\label{eq:Hephotons}
n_{4\,\Ry} \sim \frac{\dot{n}_{4\,\Ry}}{H(z=3.5)} \approx 10^{-7}\,{\rm cm^{-3}}.
\end{equation} 
Before we compare of equation (\ref{eq:Hephotons}) with (\ref{eq:nHe}), we note the it takes roughly 2-3 He ionizing photons to completely ionize He and that obscuring material around a quasar will remove half of the ionizing flux \citep{McQuinn+09}.  With these efficiency factors in mind, the comoving density of HeII ionizing photons at $z\sim 3$ is just large enough to reionize HeII.  

The amount of excess energy per HeII ionization is $16$ eV for an ionizing radiation spectral index of -1.6.  Given that there are $\approx 3$ ionizing photons per HeII reionization, the excess energy dumped into the IGM during HeII reionization is 
\begin{equation}
Q_{\rm exc,HeII} = \epsilon_{\rm exc,HeII} \frac {f_{\rm He}\Omega_B\rho_{\rmn{cr}}}{A_{\rm He}m_p} \approx  6\times 10^{-7}\,{\rm eV\,cm^{-3}},\label{eq:QHeII}
\end{equation}
where $\epsilon_{\rm exc,HeII} = 48$ eV is the average excess energy dumped per HeII ionization, $f_{\rm He} = 0.24$ is the primordial mass fraction of He, and $A_{\rm HeII} = 4$ is the atomic number of He.
To compare this to the energy dumped from TeV blazars, we start with the comoving luminosity density of blazars at $z=3.5$ is $\Lambda_B(z=3.5) \approx 3\times 10^{38}\,{\rm ergs\,s^{-1}\,Mpc^{-3}}$, which we estimate from Figure \ref{f:BLD}.  Over a Hubble time at $z=3.5$, the amount of energy dumped into the IGM by TeV blazars per comoving volume is then
\begin{equation}
 \left.Q_{\rm B}\right|_{z=3.5} =  \frac{\Lambda_B(z=3.5)}{H(z=3.5)} \approx 5 \times 10^{-7}\,{\rm eV\,cm^{-3}},
\end{equation}
which compares favorably with Equation (\ref{eq:QHeII}) and shows that as HeII reionization is being completed, the effect of blazar heating begins to be pronounced.  

For convenience, we provide a fitted third-order polynomial formula for
our estimated {\it proper} blazar heating rate for $z\le5.7$:
\begin{eqnarray}
\log_{10} \left(\frac{\dq_{\rm B}/\nbaryon}{\rm 1\,eV\,Gyr^{-1}}\right) &=& 0.0315 \left(1+z\right)^3 - 0.512 \left(1+z\right)^2 \nonumber\\ &&+ 2.27\left(1+z\right) - \log_{10} \dq_\rmn{mod}\label{eq:fit}
\end{eqnarray}
Here, $\log_{10}\dq_\rmn{mod} = \{2.38,2.08\}$ 
for the ``standard'' and ``optimistic'' (see below) blazar heating model, respectively.  The fits are
shown by the solid (gray) lines in Figure \ref{f:rates}.  Note that we have calculated the heating rate as a heating rate per baryon, i.e., $\dq/\nbaryon$, so that it is independent of the ionization fraction.

\begin{figure*}
\begin{center}
\includegraphics[width=\columnwidth]{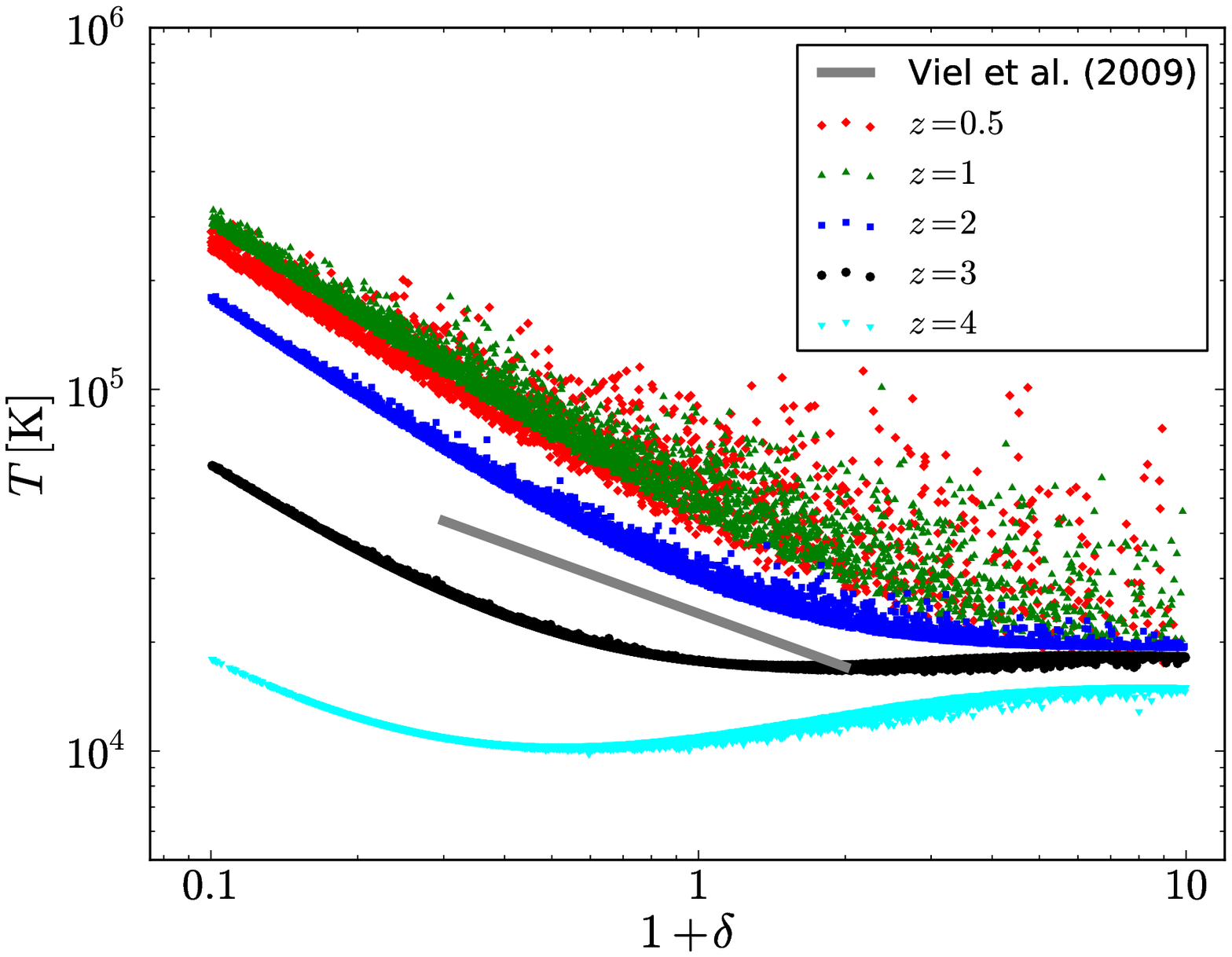}
\includegraphics[width=\columnwidth]{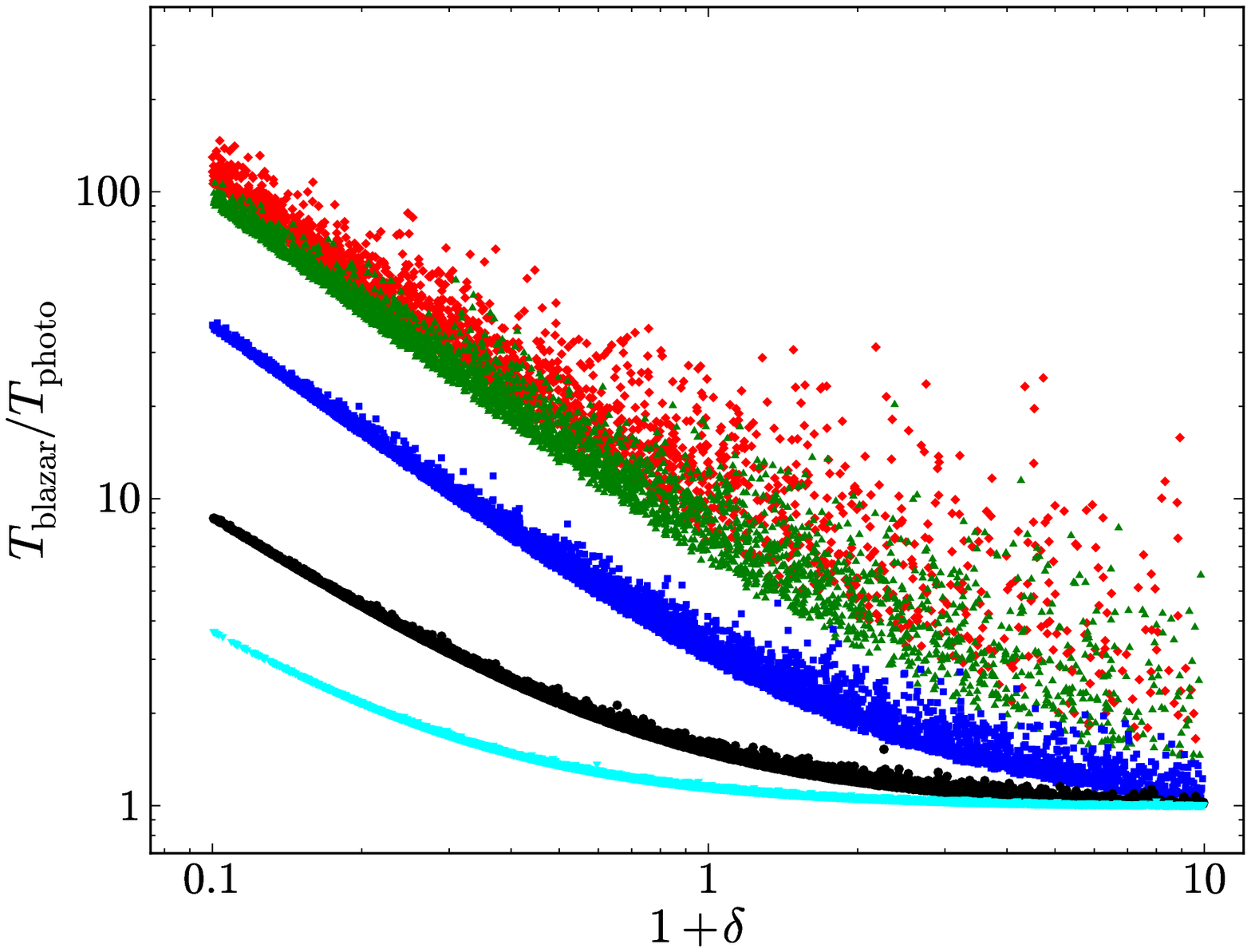}\\
\includegraphics[width=\columnwidth]{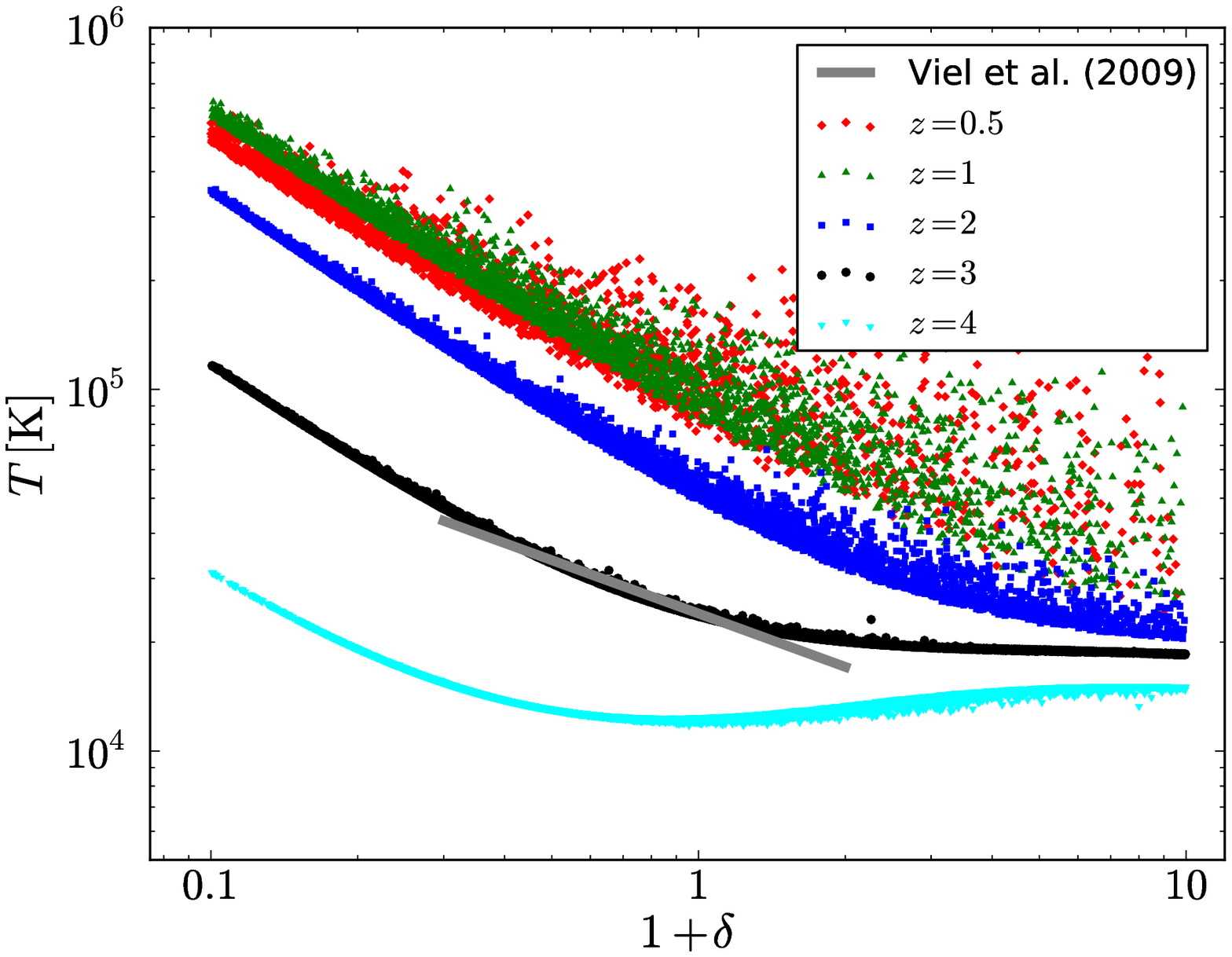}
\includegraphics[width=\columnwidth]{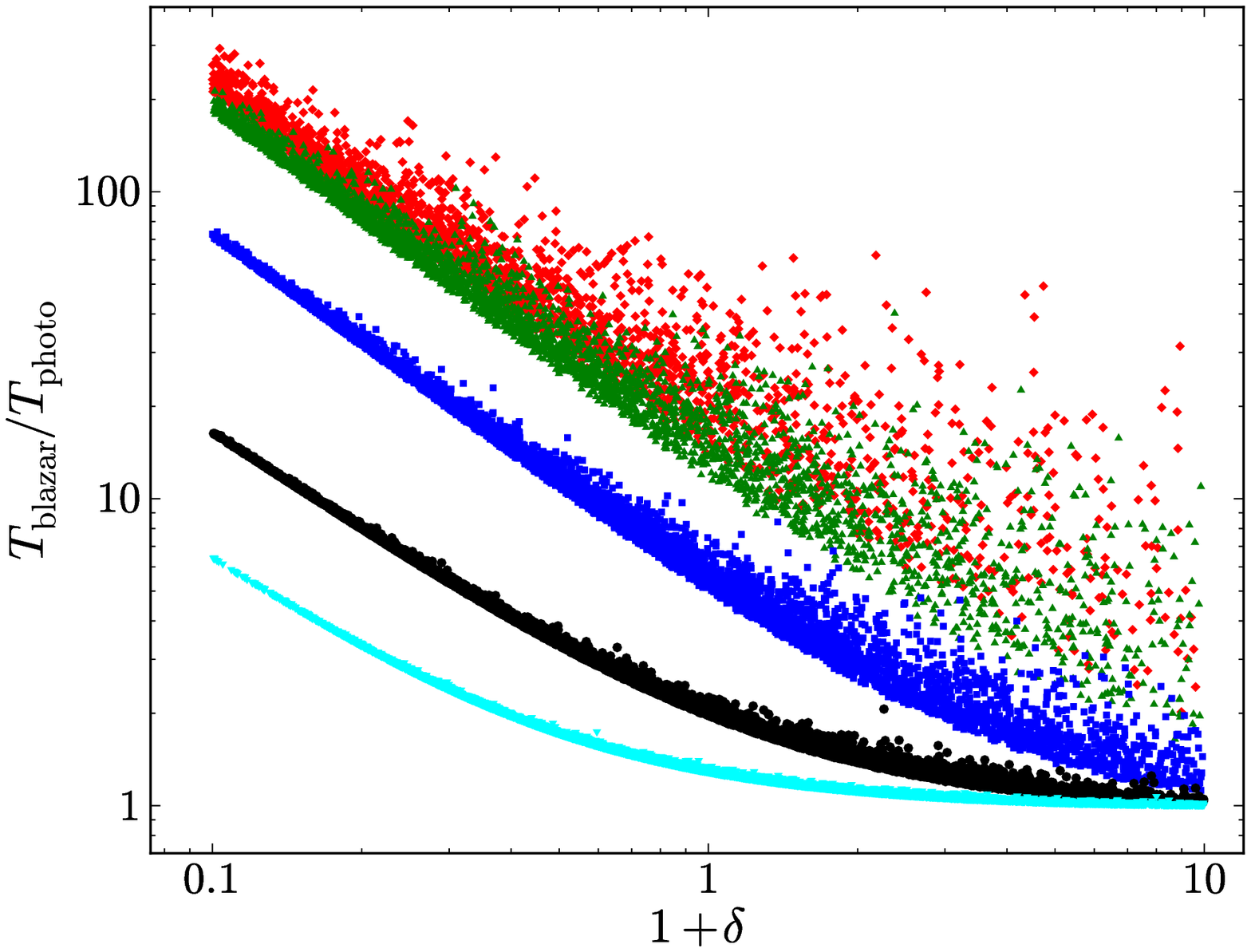}
\end{center}
\caption{Temperature-density scatter plots (left panels; same as
  Figure \ref{f:EOS}) including the effects of blazar heating for our
  standard (top) and optimistic (bottom) heating models. The right
  panels shows the ratio between including only the effects of
  photoheating ($T_{\rm photo}$) and including the effects of blazar
  heating ($T_{\rm blazar}$) for the same realization for the standard
  (top) and optimistic (bottom) blazar heating models.  Note for the
  left panels that the temperature evolution due to blazar heating is
  qualitatively different from the earlier case (Fig. \ref{f:EOS}).
  Starting at $z=4$, the temperature-density relation in the lowest
  density regions of the Universe is inverted.  This inverted
  temperature relation moves to higher densities with decreasing
  redshift until at $z=0.5$, it goes down to $\delta \approx 2$. The
  IGM temperature is also significantly hotter, with the hottest
  regions having $T > 10^5$ K. Overplotted is the observational
  determined temperature-density relation at $z=3$ (solid gray line) from
  \citep{Viel+09}. The right panels show that the temperature
  increases by a factor of $\sim$~100 (standard
heating) and $\sim$~200 (optimistic heating) for small $1+\delta$
patches of the Universe when the effects of blazar heating are taken
into account.}
   \label{f:qlf EOS}
\end{figure*}

The effect of blazar heating upon the void temperature-density
relation is shown in the left panels of Figure \ref{f:qlf EOS} for the "standard" (upper) and "optimistic" (lower) heating models.  In comparison to the
case in which blazar heating is neglected (Fig. \ref{f:EOS}), the
voids are hotter at all redshifts.  Overall, the IGM never drops below
$10^4\,\K$, with low-density regions remaining substantially hotter
than this.  By contrast, the temperature of the IGM without blazar heating does not {\it exceed} $10^4$ K (Fig. \ref{f:EOS}). In these low-density regions the temperature can exceed
$10^5\,\K$ with TeV blazar heating, almost two orders of magnitude hotter than anticipated by
photoheating and adiabatic cooling alone.  To show this result more clearly, we show in the right panels of Figure \ref{f:qlf EOS} the ratio between the temperature for a photoheating only model ($T_{\rm photo}$) and for a model including blazar heating ($T_{\rm blazar}$) for the "standard" (upper) and "optimistic" (lower) heating models.  This shows that the temperature in low-density patches in the Universe, i.e., small $1+\delta$, increase by a factor of $\sim$~100 and $\sim$~200 for the "standard" and "optimistic" heating models, respectively.

Figure \ref{f:qlf EOS} shows that blazar heating qualitatively changes the IGM temperature-density relation.  In particular, the IGM
temperature-density relation is inverted for $\delta\lesssim0$.  This compares
favorably with recent evidence for just such an inverted temperature-density relation \citep{Bolton+08,Viel+09}.  Notably, in the lower panel of Figure \ref{f:qlf EOS}
we plot as a solid (gray) line the temperature-density relation
$T = 2.4\times 10^4\,(1+\delta)^{-0.49}\,\K$, inferred at $z=3$
empirically by \cite{Viel+09}.  Here to produce a better match to the empirical measurements of \citet{Viel+09}, we introduce the ``optimistic'' heating model, i.e., we choose a value of the remaining coefficient of the sky incompleteness correction for TeV blazars of $\eta_\rmn{sys}=1.6$, which includes variations of the TeV blazar redshift evolution, additional TeV sources that may contribute to the plasma instability heating, and potentially spectral variability. As a result, the local heating rate is enhanced over the "standard" heating rate by a factor of two or $\left.\dq\right|_{z=0} \approx 1.4 \times 10^{-7} \eV\,{\rm cm^{-3}Gyr^{-1}}$. The match produced between the empirical relation of \citet{Viel+09} and our ``optimistic'' TeV blazar model  is striking in that it matches the slope in the absence of any tuning as well as the normalization within the uncertainties of our estimated incompleteness correction (see Figure~\ref{f:BLD}).  This inversion of the temperature-density relation is difficult to reproduce using HeII reionization alone
\citep{McQuinn+09,Bolton+09} though is a natural consequence of TeV
blazar emission.

Energetically, the magnitude of the impact from TeV blazars in
low-density regions is somewhat surprising.  The radiative
output from stars and quasars far outstrips that from VHEGR
sources, yet in practice the heating rate from blazars is much
larger.  This is because the photoheating rate is ultimately limited
by the HII recombination rate.  Indeed, this is precisely the property
invoked to show that the effect of the epochs of H and HeII
reionization is washed out at low redshift, unless these occurred
recently \citep{Hui03}.  

By contrast no such limitation exists for TeV blazars, paving the way for them to dominate the heating of the low-redshift Universe, a point that we have made already in the Introduction.  Instead, the VHEGR emission from blazars is deposited in the IGM with
order unity efficiency.  The rate of heating in this case
depends linearly on the radiative output of these TeV blazars, and is
independent of the atomic physics of the IGM.  In addition these
sources have a constant heating rate per unit volume (whereas
photoheating heats per unit mass).  Hence, the effect of 
blazar heating, which is already pronounced, is amplified relative to
photoheating in low-density regions, leading to an inverted temperature-density relation. Therefore, not only is the memory of photoheating is erased in
low-density regions, but it is overwritten by the record of blazar heating. 

Following this work, we have recently completed a study of the effects
of blazar heating in a more detailed hydrodynamic model of structure
formation, reported in a follow-up paper \citep{Puchwein+2011}. Using
the blazar heating prescription given by equation (\ref{eq:fit}), we 
show that the optical depth of the \Lya\ forest is reproduced using 
a H photoionization rate of $\Gamma_{\rm HI} \approx 5\times 10^{-13}\,{\rm s}^{-1}$ or 
equivalently using the inferred ionizing background from \citep{FG+09}.
We also confirm that the low density IGM again possesses an inverted
temperature-density relation (as shown in Figure \ref{f:qlf EOS}).  In
addition, a detailed comparison between the results of our numerical
calculations and observations show that a blazar heated universe
matches the one- and two- points statistics of the high redshift
Ly$\alpha$ forest, but also matches the line width distribution.  This
excellent agreement was achieved using the best estimate on the
evolution of the photoionizing background (without tuning) and is due
to the excess energy injection into the low density IGM.

\subsection{Implications for the local \Lya forest} \label{sec:Lyalocal}

While TeV blazar heating can with significant success reproduce some of
the peculiar properties of recent high-$z$ \Lya observations, 
the most dramatic departures from the IGM thermal history in the
absence of blazar heating occur at $z\lesssim 1$.  This is a result of
the relatively recent nature of quasar activity and the cumulative effect of blazar heating.
Thus we might anticipate dramatic consequences for the local \Lya
forest, potentially in conflict with studies of nearby \Lya clouds
\citep[e.g.,][]{PentonI+2000,PentonII+2000,Penton+2002,Penton+2004,Dave_etal:10}.
However, this does not occur due to the physical nature of
the clouds that produce the local \Lya forest.

The structures responsible for the local \Lya forest are almost
certainly associated with significant overdensities as suggested by large-scale hydrodynamic computations
\citep[e.g.,][]{Dave_etal:10}\footnote{We note in passing, that for
  $z\gtrsim2$, the same simulations imply that \Lya forest contains
  significant contributions from regions with $1+\delta\sim1$.}.
These simulated \Lya forest calculations provide a means to relate the empirically inferred
HI columns to the properties of the dynamical structures responsible
for the \Lya clouds.  For HI columns of
$10^{13}$--$10^{14}\,\cm^{-2}$, the lowest values for which \Lya
measurements exist, the simulations find that 
the local \Lya forest is produced primarily by the intergalactic
filaments, corresponding to $1+\delta\gtrsim5$--$500$ 
\citep[e.g., see the bottom-left panel of Figure 8 and Figure 9 of ][]{Dave_etal:10}.

We emphasize this point with the following order-of-magnitude estimate.  An upper limit upon cloud sizes can be obtained
using the line widths, $b$, translating into a proper size of
$1.4 (b/10^2\,\km\,\s^{-1})\,\Mpc$  
\citep{PentonII+2000}.  A lower limit upon the local ionizing
background can be inferred from the optical depth of VHEGRs, giving $\sim10^{-6}\,\erg\,\s^{-1}\,\cm^{-2}\,\Sr^{-1}$
\citep{Ahar_etal:06}.  Employing ionization balance to set the HI
fraction, for $z<1$ this gives the column densities
$\lesssim3\times10^{10} (1+\delta)^2 (1+z)^{6-\zeta}
(T/10^4\,\K)^{-0.65}$ for $T<10^5\,\K$.  
From this it is clear that the nearby \Lya clouds must correspond to
regions with $1+\delta\gg1$.\footnote{Note, however, that since the
  column density is $\propto(1+z)^6$ at $z>1$, we would nevertheless expect
  low-density regions to contribute significantly to the high-$z$ \Lya
  forest, as has indeed been found to be the case.}

At these overdensities, the impact of blazar heating on their thermal history is modest at best.
By $1+\delta=5$ the blazars change the IGM temperature
by a factor of 2, raising it to roughly $4\times10^4\,\K$, comparable
to the temperatures typical of the local \Lya clouds
\citep{Dave_etal:10}.  At $1+\delta>50$ blazar heating is negligible
in comparison to photoionization and shock-heating.  Thus, because low-$z$ \Lya absorbers are biased toward high-density regions, they serve a poor probes of low-density regions and are unaffected by the effects of TeV blazars.

The inclusion of blazar heating into these large-scale hydrodynamic
calculations is an important next step. However, we do not expect a significant
impact on the local \Lya forest.  A more important impact of TeV
blazar heating would be on the formation of collapsed structures, which is a topic that we will explore in Paper III.

\subsection{Limits on the direct detection of the hot IGM}

As low-density regions are immune to local \Lya probes, we now turn to
the question whether there are methods of directly
inferring the presence of such high temperatures in these
regions. First, we compute the mean Comptonization of the CMB due
to blazar heating,
\begin{equation}
  \label{eq:mean_y}
  \langle y \rangle = \sigma_T \int_0^{l_\rmn{HI}} d D
  \frac{\langle n_e k\, (T_e - T_\rmn{CMB}) \rangle}{m_e c^2},
\end{equation}
where $n_e$ is the free electron density.  Here, $n_e$ is the physical
density of free electrons and we integrate along the proper distance
of the photons back to recombination, $d D = -c\, da \dot{a}^{-1} =
-c\,da (a H)^{-1}$ where the minus sign arises due to the choice of
coordinates which are centered on the observer. Performing the
integral with our temperature evolutions at mean density
(i.e. neglecting gravitational heating by formation shocks) yields
mean Comptonization values of $\langle y \rangle =
\{1.4,1.9,2.5\}\times 10^{-7}$ for our models without blazar heating
and those with standard and optimistic blazar heating. To date the
best limits on the mean Comptonization come from the COsmic Background
Explorer Far-InfraRed Absolute Spectrophotometer experiment (COBE
FIRAS) which measure the difference between the CMB and a perfect
black-body spectrum \citep{Fixsen+1996}. Their upper limit of $|y| <
1.5\times 10^{-5}$ (95\% confidence level) is perfectly consistent
with our inferred mean Comptonizations. With current technology,
  it appears to be quite feasible to measure the deviation of the CMB
  spectrum from a perfect blackbody form with an accuracy and
  precision of 1 ppm yielding constraints on the cosmic $y$-parameter
  at the level of $10^{-7}$ and provide a spectrum of the anisotropy
  to 10\% \citep{Fixsen+2002}.  We note, however, that the mean
Comptonization is expected to be dominated by gravitational heating
which leads to values of $\langle y \rangle = 2.6\times 10^{-6}$
inferred from cosmological simulations by \citet{Springel+2001}. The
mean temperature of these cosmological simulations of $\langle T_0
\rangle = 0.3\,\keV$ suggests that the signal is dominated by
collapsed galaxy groups and unvirialized infall regions onto galaxy
clusters that constitute the hot component of the warm-hot
intergalactic medium. Hence, in order to measure the blazar
  heating signal in the mean Comptonization of the CMB, the
  fluctuating part due to gravitational heating would have to be
  subtracted first. Since galaxy groups dominate the fluctuation power
  on angular scales of $<5'$, the measurement of the deviation of the
  CMB spectrum would have to be performed on these small angular scales which
  appears to be difficult.

Second, we estimate the emission of the IGM due to free-free
  emission (bremsstrahlung) and synchrotron emission.  Turning to the
question of synchrotron emission, we estimate the synchrotron
frequency using a maximal IGMF strength of $B\sim 10^{-9}$ G.  At
temperatures of $\sim 10^4-10^5$ K, the electrons in the IGM are
nonrelativistic.  Hence the synchrotron frequency is roughly the
Larmor frequency, $\omega_{\rm sync} \approx \omega_{\rm G} = eB/ m_e c \approx 2\times 10^{-2}\left(B/10^{-9}\,{\rm G}\right)\,{\rm s}^{-1}$.  
These low frequencies are well below the plasma frequency of the IGM, 
$\omega_p = \sqrt{4\pi e^2 n_e/m_e} \approx 25$ Hz and, hence, will be
absorbed by the IGM. Free-free emission from the IGM can cause a distortion in the CMB at low frequencies \citep{Bartlett+91}.  As the free-free emissivity scales like $\propto n_e^2 T^{-0.35}$ for fully ionized primordial gas, it scales weakly with temperature but strongly with clumpiness.  As a result, overdense haloes dominate free-free emission, with the contribution from low density IGM being 1-3 orders of magnitude smaller \citep{Oh99,Cooray+04,Ponente+11}.  The effect of blazar heating will further suppress this free-free emission (albeit mildly) from the IGM compared to that of overdense halos.  Finally, the best constraints on this free-free optical depth of $\tau_{\rm ff}  < 1.9\times 10^{-5}$ \citep{Bersanelli+94} is still too large compared to optimistic models of free-free emission from overdense haloes by at least an order of magnitude and by $\approx 3$ orders of magnitude from the smooth IGM \citep{Cooray+04,Ponente+11}.

\section{Measuring the High-Energy Luminosity of Blazars}\label{sec:measuring blazars}

Thus far, we have concerned ourselves with the impact of TeV blazars
on the Universe at large.  In this section, we invert this argument
and discuss how the Universe at large can be used as a probe of the physics of VHEGR photons and the
global properties of TeV blazars.  Namely, the estimated heating rates
used in this paper -- standard and optimistic -- suffer from a number
of uncertainties, including statistical fluctuations in the number of
presently observed VHEGR sources, the redshift evolution of the TeV blazar luminosity density, and
the detailed form of the VHEGR and EBL spectra.  However, due to the
high efficiency with which the VHEGR luminosity of
blazars is converted to heat within cosmic voids, the temperature
history of the voids themselves provides a way to determine the
cumulative TeV blazar luminosity density empirically.

This is possible due to a number of fortuitous properties of voids
and TeV blazar heating.  First, the
high efficiency with which the VHEGR emission of blazars is converted to
heat implies that even in voids the Universe acts as a calorimeter.
Second, blazar heating
dominates the thermal history of voids for $z\lesssim4$ (see, e.g.,
Figure \ref{f:rates}) so that this calorimeter is uncontaminated.  
We caution that HeII reionization complicates matters somewhat, though in our model
makes a comparatively small correction to the temperature evolution of
low-density regions (see Figure \ref{f:thermal history}).  However,
this depends upon the particular manner in which HeII reionization
occurred, and could in principle provide a somewhat larger
contribution in inhomogeneous reionization scenarios \citep[see
  Section 3 of][ for a detailed discussion]{Furlanetto08}.
Finally, within voids the adiabatic losses to Hubble expansion are
very accurately modeled in the linear regime, substantially simplifying
the interpretation of void temperature histories.

Perhaps the greatest uncertainty is the physics of the mechanism which converts pair beams from TeV blazars to thermal energy in the IGM.  In the context of the "oblique" instability, what is unclear is the luminosity cutoff of TeV blazars probed in this manner,
which arises from the competition between inverse-Compton and plasma
processes.  If this cutoff may be conservatively assumed to lie at an
isotropic-equivalent $E L_E \simeq 10^{42}\,\erg\,\s^{-1}$, 
corresponding to a true luminosity 2--3 orders of magnitude lower due
to the presumed jet beaming, then this is dimmer than all but two
of the TeV blazars known.  Assuming that the TeV blazar luminosity function
follows that of quasars, the luminosity density is dominated by
sources near the break luminosity, and thus as long as the
lower-luminosity cutoff is sufficiently low (below that of the break
luminosity) it may be neglected.  For the TeV blazars listed in Table \ref{tab:TeVsources}, this break occurs near
$3\times10^{44}\,\erg\,\s^{-1}$, and thus is well above the relevant
cutoff.  Measurements of the thermal history of voids directly
corresponds to the TeV blazar bolometric luminosity evolution.  In this manner, the 
thermal history of cosmic voids provides an analogous argument to that by \citet{Solt:82} for
  determining the blazar luminosity density, $\Lambda_B(z)$.  Such a constraint on the 
history of 
$\Lambda_B(z)$ can be used to study the history of accretion in
the Universe and the jet forming efficiency by comparing it to the
history quasar or active galaxy luminosity density.  While probes of the low-density IGM at low redshift are sparse, the situation at $z\sim 2-4$ is much more hopeful as we have discussed in Section \ref{sec:IGM blazar heating}.  Thus, precision measurements of the \Lya forest at $z\sim 2-4$ \citep[e.g.][]{Viel+09} alongside detailed studies of HeII reionization \citep[e.g.]{McQuinn+09} offer the best constraint both on the fate of VHEGR photons and the evolution of TeV blazars, i.e., the blazar luminosity density.

\section{Conclusions}\label{sec:conclusions}

In this work, we have explored the effect of TeV blazar heating on the
thermal history of the IGM. We have argued that VHEGRs
that are sufficiently hard to pair produce off of the EBL, will
inevitably dump the majority of their energy into the IGM via plasma
instabilities.  By collating the nearby 28 TeV blazars with firm
spectral measurements, we have determined the local observed heating
rate after correcting for the various selection effects using \Fermi
observations of TeV blazars.  This local heating rate of
$\dq = 7\times 10^{-8}\,{\rm eV\,cm^{-3}\,Gyr^{-1}}$, which we call the
standard model, can be extended to higher redshift by normalizing it
to the \citet{Hopkins+07} quasar luminosity density.  This follows
from the important result of Paper I, which shows that the local
observed blazar luminosity function is well in line with the local
quasar luminosity function corrected by a factor of $\approx
10^{-3}$.  This TeV blazar heating should be relatively
homogeneous at all redshifts $z\lesssim 4$ with greater spatial variations for at higher $z$, approaching order unity at $z\sim 6$.

This redshift dependent blazar heating is substantial and is larger than the photoheating rate
by a factor of 15 (standard) - 30 (optimistic) after He reionization
for a $\delta = 0$ patch of the Universe.  Using a
simple one-zone model of the IGM, we demonstrate that the effect of
including TeV blazar heating versus not including TeV blazar heating
leads to qualitative and quantitative changes in the thermal history
of the IGM. First, the injection of heat into the IGM by blazars
substantially increases the temperature of the IGM.   In the case with
blazar heating, the temperature of the IGM stays above $10^4$ K, with
some regions approaching $10^6$ K, whereas without blazar heating, the temperature of the IGM
tends to stay below $10^4$ K.\footnote{Again, we have ignored the effects of gravitational (shock) heating which is important at densities above mean density.}  Second, the even volumetric heating
rate of blazars impacts the thermal history of low-density regions
much more strongly than higher density regions.  Low-density regions
are substantially hotter as a result with temperatures in excess of
$\gtrsim 10^5$ K.   Higher density regions on the other hand are not
heated as much by blazars.  This naturally produces an
inverted temperature-density relation which matches the empirical results of
\citet{Viel+09} if we increase the amount of blazar heating, i.e., the
optimistic model, to
$\dq = 1.4\times 10^{-7}\,{\rm eV\,cm^{-3}\,Gyr^{-1}}$. It also provides
an encouraging endorsement of our model as such inverted
temperature-density relations are difficult to produce in standard
reionization histories.  We have demonstrated these salient points
more explicitly in a follow up paper \citep{Puchwein+2011}, where we
calculate the effect of blazar heating in a hydrodynamic realization
of the universe.  We show that the comparison between a blazar heated
universe and observation of the high redshift Ly$\alpha$ forest gives
excellent quantitative agreement.

As our model predicts a substantially hotter low-density IGM that
standard models predict, we then proceeded to investigate either if
this model breaks current constraints on the local temperature of the
IGM or can be directly measured. Unfortunately, the local \Lya forest
is an ineffective probe of this environment compared to the high-$z$
\Lya forest as the regions that give rise to the local \Lya forest are
dense regions that remain relatively unaffected by the effects of
blazar heating.  Other means of directly probing this hot IGM
via Comptonization of the CMB and free-free emission
emission are also unlikely.

Finally, we note that the thermodynamics of the IGM can be used as a
calorimeter for VHEGR emission of blazars in the Universe.  Namely,
because the low-density IGM is so sensitive to the total amount of
energy dumped into it, which is dominated by TeV blazars, we argue
that the thermal history of the low-density IGM can be used to measure
the total energy output in VHEGRs over cosmic time.
In principle, this would allow a determination of the blazar luminosity density as a
function of redshift, as well as constrain the history and physics of
accretion onto supermassive black holes, i.e., rates of radiative
versus radiative inefficient accretion and jet formation efficiency.
However, the contaminating effects of He II reionization would have to
be explored in detail before such physics can be elucidated.

\acknowledgements We thank Tom Abel, Marco Ajello, Marcelo Alvarez,
Arif Babul, Roger Blandford, James Bolton, Mike Boylan-Kolchin, Luigi
Costamante, Andrei Gruzinov, Peter Goldreich, Martin Haehnelt, Andrey Kravtsov, Ue-li
Pen, Ewald Puchwein, Volker Springel, Chris Thompson, Matteo Viel,
Marc Voit, and Risa Wechsler for useful discussions.  We also thank the referee, David 
Weinberg, for a thorough reading of the manuscript and for his constructive comments.  
We are indebted
to Peng Oh for his encouragement and useful suggestions. We thank
Steve Furlanetto for kindly providing technical expertise. These
computations were performed on the Sunnyvale cluster at CITA.
A.E.B. and P.C. are supported by CITA. A.E.B. gratefully acknowledges
the support of the Beatrice D. Tremaine Fellowship.  C.P. gratefully
acknowledges financial support of the Klaus Tschira Foundation and
would furthermore like to thank KITP for their hospitality during the
galaxy cluster workshop.  This research was supported in part by the
National Science Foundation under Grant No. NSF PHY05-51164.

\begin{appendix}
\section{Defining the Number of Bright Blazars}

Here we provide a more detailed discussion of how ``likely'' it is
for a patch of the Universe at a given redshift to experience a large
deviation from the average TeV blazar heating rate.  We will pursue
this primarily by attempting to compute the ``number of blazars a patch
sees'' as a function of redshift, ${\mathcal N}_B(z)$.  This is, however, a
poorly defined quantity, the primary 
difficulty being the determination of which objects to count.  In
principle, we would like to take a census of those sources
``responsible for the bulk of the heating''; in practice this is an
ambiguous definition.  Thus, here we will describe and contrast a
number of possible definitions.  In the process, we will also
elucidate which objects dominate the heating.

\begin{figure}
\begin{center}
\includegraphics[width=0.95\columnwidth]{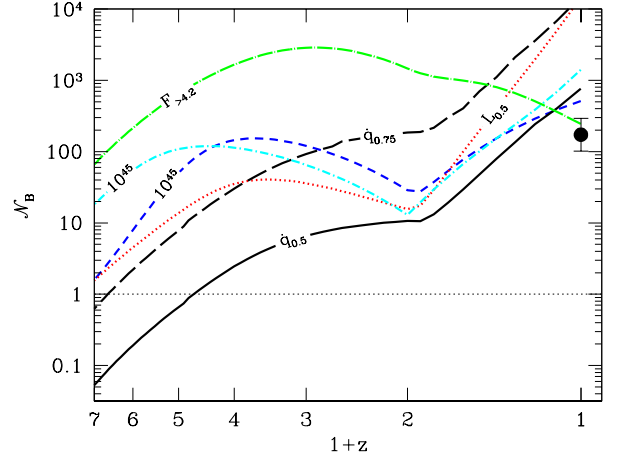}
\end{center}
\caption{Same as Figure \ref{f:numBlazars} with the addition of the
  cyan, short-dash-dotted line, showing the instantaneous number of blazars
  within a comoving volume of radius $(1+z)\bDpp$, and the green,
  long-dash-dotted line, showing the number of blazars seen at a particular
  redshift above a flux limit of
  $4\times10^{-12}\,\erg\,\cm^{-2}\,\s^{-1}$, roughly that inferred
  from the very-high energy gamma-ray observations.
  The remaining lines are as defined in Figure \ref{f:numBlazars}:
  The blue, short-dashed line shows the instantaneous number
  of blazars with intrinsic isotropic-equivalent luminosities above
  $10^{45}\,\erg\,\s^{-1}$ within the $\bar{\tau}=1$ surface, the red,
  dotted line shows the number of blazars above the median luminosity
  at each redshift within the $\bar{\tau}=1$ surface, and the 
  black, solid (long-dashed) line shows the number of blazars with individual 
  heating rates that exceed that above which half (0.75 times) of the
  heating is produced.  For reference, our estimate of the sky-completeness
  corrected number of TeV blazars {\em observed in the TeV} is shown
  by the black circle, with the error bars denoting the Poisson
  uncertainty only.  Note that for all calculations we have assumed
  $L_M=2\times10^{46}\,\erg\,\s^{-1}$ and adopted $\alpha=3$.
}
\label{f:NBs}
\end{figure}

\subsection{Preliminary Definitions and Assumptions}
In the interest of completeness we will forgo the assumption that the
heating is local, and perform the appropriate cosmological
calculations.  However, since we ultimately would like to identify
${\mathcal N}_B$ as a function of observer redshift, $z_o$, we require
observed-dependent definitions of the standard compliment of
cosmological distances.  Specifically, we use the following
generalizations of the standard proper, angular diameter,
and luminosity distances,
\begin{equation}
\begin{aligned}
D_P(z;z_o) &= D_P(z)-D_P(z_o)\\
D_A(z;z_o) &= \frac{D_C(z)-D_C(z_o)}{1+z}\\
D_L(z;z_o) &= (1+z)\frac{D_C(z)-D_C(z_o)}{(1+z_o)^2}\,,
\end{aligned}
\end{equation}
where $D_C(z)$ and $D_P(z)$ correspond to the $z=0$ comoving and
proper distances, respectively.  Note that these reduce to their
expected expressions when $z_o=0$.

While we have already defined the mean free path of high-energy gamma
rays in Equation (\ref{eq:Dpp}), we also require a definition of the
optical depth that a gamma ray that originates at $z$ with
{\em observed} energy $E$ at $z_o$ accrues during its propagation:
\begin{equation}
\tau(E,z;z_o)
=
\int_{z_o}^z \frac{1}{D_{pp}\left[(1+z)E/(1+z_o),z\right]} \frac{d D_P}{dz} dz\,.
\label{eq:tau}
\end{equation}
As with $\tau(E,z)$, this differs from the definition of $\tau_E(E,z)$
at $z_o=0$ given in Paper I, where there we set $E$ to the 
{\em emitted} energy of the gamma ray.

We define the flux, $F(E_m,E_M)$, to be that integrated between a
given energy range, $E_m$ to $E_M$ (usually 100 GeV to 10 TeV),
\begin{equation}
F(E_m,E_M) = \int_{E_m}^{E_M} dE\, F_E
=f_0 E_0^{\alpha}\, \int_{E_m}^{E_M} E^{1-\alpha} dE \,,
\end{equation}
where the observed photon number flux is given by Equation (\ref{eq:spectra}).
For a source at redshift $z$, this corresponds to an absorption
corrected flux of
\begin{equation}
\bar{F}(z;z_o,E_m,E_M)
=
\int_{E_m}^{E_M} dE F_E e^{-\tau(E,z;z_o)}\,,
\end{equation}
and therefore an intrinsic luminosity between energies
$E_m(1+z)/(1+z_o)$ and $E_M(1+z)/(1+z_o)$ of
$L(z;z_o,E_m',E_M')=4\pi D_L(z;z_o)^2\bar{F}(z;z_o,E_m,E_M)$ where $E_m'$
and $E_M'$ are the energies bounding the relevant range at $z$.  Since
we would like to compare luminosities within a band across redshifts
(i.e., keep $E_m$ and $E_M$ fixed), we must correct for the fixed
spectral shift induced by the different redshifts.  Assuming that the
spectrum is a power law (as we shall do in all cases here), this
implies an additional redshift factor: 
\begin{equation}
\begin{aligned}
L(z;z_o,E_m,E_M)
&=
4\pi \left(\frac{1+z}{1+z_o}\right)^{\alpha-2} D_L^2(z;z_o) \,\bar{F}(z;z_o,E_m,E_M)\\
&=
4\pi \left(\frac{1+z}{1+z_o}\right)^{\alpha-2} D_L^2(z;z_o)
\,\e^{-\bar{\tau}} \,F(E_m,E_M)\,,
\end{aligned}
\label{eq:L}
\end{equation}
where $\bar{\tau}$ is the spectrally average optical depth:
\begin{equation}
\bar{\tau}(z;z_o)
=
-\ln\left[
\frac{
\int_{E_m}^{E_M} E^{1-\alpha} e^{-\tau(E,z;z_o)}dE
}{
\int_{E_m}^{E_M} E^{1-\alpha} dE
}
\right]
\end{equation}
Note that this is closely related to $\bDpp(z_o)$.

Finally, we will adopt the form of the physical density of TeV
blazars, $\tBLF(L,z)$, described in the main text assuming a cutoff at
$L_M=2\times10^{46}\,\erg\,\s^{-1}$, consistent with physical models
of high-energy gamma ray blazars \citep[see,
  e.g.,][]{Ghisellini+2009}.  Associated with this we have a
generalized number density,
\begin{equation}
\tilde{\Phi}_B(z;L_m,L_M) = \int_{\log_{10}L_m}^{\log_{10}L_M} \tBLF(z,L) \dlL
\end{equation}
and luminosity density,
\begin{equation}
\tilde{\Lambda}_B(z;L_m,L_M) = \int_{\log_{10}L_m}^{\log_{10}L_M} L \tBLF(z,L) \dlL\,,
\end{equation}
defined within a given intrinsic luminosity range.

\subsection{Possible Definitions of ${\mathcal N}_B$}
We now present various ways to define the number of high-energy
gamma-ray blazars that are relevant for heating.  These are compared
in Figure \ref{f:NBs}, which supplements Figure \ref{f:numBlazars} with
additional approximations for ${\mathcal N}_B$.

\subsubsection{Local number within a mean free path}
Our first definition is the simplest one can imagine; choose an
intrinsic luminosity range and use the local density to determine the
number within a volume defined by the spectrally-average mean free
path.  That is,
\begin{equation}
{\mathcal N}_{B,I}(z_o;L_m)
=
\frac{4\pi}{3} \bDpp^3(z_o) \tilde{\Phi}_B(z_o;L_m,L_M)\,.
\label{eq:NI}
\end{equation}
This is an approximation of the number of sources with intrinsic
luminosities in the specified range within an approximation of a
single optical depth.  This is shown for $L_m=10^{45}\,\erg\,\s^{-1}$
by the cyan dash-dotted line in Figure \ref{f:NBs}.  Typically, where the
blazar population is rapidly evolving it tends to be a poor estimate
of the number of sources, overestimating this number by nearly an
order of magnitude at $z\gtrsim3$.

\subsubsection{Number within $\bar{\tau}=1$}
As long as $\tBLF(z,L)$ evolves slowly and $\bDpp\ll c/H_0$ it is
unnecessary to perform the relevant redshift integral to get the volume
element.  However, this is not always the case, and thus a more
accurate estimate is obtained by integrating the blazar number density
within the volume specified by unit optical depth.  That is, 
\begin{equation}
{\mathcal N}_{B,II}(z_o;L_m)
=
\int_{z_o}^{z_1} 4\pi D_A^2(z;z_o) \frac{dD_P}{dz}
\tilde{\Phi}_B(z;L_m,L_M) \,dz\,,
\label{eq:NII}
\end{equation}
where $z_1$ is defined implicitly by $\bar{\tau}(z_1;z_o)=1$.

For this definition, we may choose $L_m$ in a variety of ways.  In
Figure \ref{f:NBs} we show two in particular: that from setting
$L_m=10^{45}\,\erg\,\s^{-1}$ (blue short-dashed line), which may be directly
compared with the case shown for ${\mathcal N}_{B,I}$, and setting $L_m=L_{0.5}$
as defined by Equation (\ref{eq:L0.5}), i.e., the luminosity-weighted
median luminosity, above which sources produce half of the total
local luminosity density (red dotted line).  For $z\lesssim3$
this approximation for ${\mathcal N}_B$ gives similar results to
${\mathcal N}_{B,I}$ for a
fixed $L_m$.  At high redshifts, where $\tBLF$ is rapidly decreasing,
the two diverge substantially.  The difference between the
${\mathcal N}_{B,II}(z;L_{0.5})$ and the previous two is more striking, and a
consequence for the assumed evolving luminosity distribution of
blazars, which is clearly important.

\subsubsection{Number of objects above a flux limit}
While a fixed intrinsic luminosity limit may be useful conceptually,
even idealized surveys do not directly probe such a population.
Instead, most surveys are flux limited, and thus we also define a
flux-limited definition of ${\mathcal N}_B$.  In this case we set the minimum
luminosity via
\begin{equation}
L_m(z;z_o,F_m) = 4\pi \left(\frac{1+z}{1+z_o}\right)^{\alpha-2}
D_L^2(z;z_o) e^{\bar{\tau}(z;z_o)} F_m \,,
\end{equation}
where $F_m$ is a fixed flux limit, from which we obtain
\begin{multline}
{\mathcal N}_{B,III}(z_o;F_m)
=
\int_{z_o}^{\infty} 4\pi D_A^2(z;z_o) \frac{dD_P}{dz}\\
\times
\tilde{\Phi}_B\left[z;L_m(z;z_o,F_m),L_M\right] \,dz\,,
\label{eq:NIII}
\end{multline}
This corresponds to the number of objects a flux-limited survey (in
the $100\,\GeV$--$10\,\TeV$ band) performed by an observer at redshift
$z_o$ would detect.  As such, it is the most directly comparable to
the number of TeV blazars that have been observed.  This is shown for
flux limit comparable to that inferred for the TeV sample in Table
\ref{tab:TeVsources}, $4.19\times10^{-12}\,\erg\,\cm^{-2}\,\s^{-1}$ in
Figure \ref{f:NBs} by the green long-dash-dotted line.  Note that the number
of objects found at $z=0$ corresponds nicely to the number observed after correcting for incompleteness of present TeV surveys.
Of course, this is by construction since $\tBLF$ was obtained from the
observed population.  Nevertheless, it is striking that many more
objects would have been observed above this flux limit during earlier
epochs, vastly exceeding any of the preceding approximations of
${\mathcal N}_B$.  However, it does not necessarily follow that all of these
sources will have contributed substantially to the heating rate.

\subsubsection{Number of objects above a fractional heating rate imposed heating rate limit} \label{sec:Nqdot}

The most relevant approximation for ${\mathcal N}_B$, and the one we adopt as our
primary definition, is set by the heating rates directly.  The general
idea is to do what we do naturally at Earth: arrange all of the
sources visible by an observer at $z_o$ by the local heating rate they
induce, from largest to smallest, and count until a fixed fraction of
the total heating rate is reached.  That is, set ${\mathcal N}_B$ to be the minimum
number of sources (on average) required to produce a given fraction of
the total heating rate.  
Given its direct connection with the heating rate, this provides the most
natural definition of the number of sources ``responsible for the bulk
of the heating.''
To do this, however, we first must explicitly
define the heating rate in terms of the appropriate functions.

Given a fixed spectrum, there is a linear relationship between the
local flux and the heating rate, defined by Equation (\ref{eq:qdot}):
\begin{equation}
\begin{aligned}
\dot{q}
&=
\int_{E_m}^{E_M} \frac{F_E e^{-\tau(E,z;z_o)}}{\Dpp\left[E(1+z)/(1+z_o),z\right]} dE
=
\chi(z;z_o) F(E_m,E_M)\\
&=
\chi(z;z_o) \left(\frac{1+z}{1+z_o}\right)^{2-\alpha}
\frac{L}{4\pi D_L^2(z;z_o)}\,,
\end{aligned}
\end{equation}
where
\begin{equation}
\chi(z;z_o)
\equiv
\int_{E_m}^{E_M} \frac{
  E^{1-\alpha} e^{-\tau(E,z;z_o)}
}{
  \Dpp\left[E(1+z)/(1+z_o),z\right]
}
dE 
\bigg/
\int_{E_m}^{E_M} E^{1-\alpha} dE
\,,
\end{equation}
is a function of the shape of the spectrum and the redshifts, similar
to $\bar{\tau}(z;z_o)$.  Thus, a given heating rate defines an
intrinsic luminosity limit for a source at a given redshift:
\begin{equation}
L_m(z;z_o,\dot{q}_m) = 4\pi \left(\frac{1+z}{1+z_o}\right)^{\alpha-2}
\frac{D_L^2(z;z_o)}{\chi(z;z_o)} \dot{q}_m \,.
\label{eq:Lmqddef}
\end{equation}
With this, we may compute the heating rate as a function of
$\dot{q}_m$, i.e., the heating rate associated with sources that
produce a local heating larger than some limit:
\begin{equation}
\begin{aligned}
\dot{Q}(z_o;\dot{q}_m)
&=
\int_{z_o}^{\infty}
4\pi D_A^2(z;z_o) \frac{dD_P}{dz}\\
&\qquad\qquad\times
\int_{\log_{10}L_m(z;z_o,\dot{q}_m)}^{\log_{10}L_M}\hspace{-0.7cm} 
 \dot{q} \,\tBLF(z,L)\,\dlL \,dz\\
&=
\int_{z_o}^{\infty}
4\pi D_A^2(z;z_o) \frac{dD_P}{dz}
\left(\frac{1+z}{1+z_o}\right)^{2-\alpha}\\
&\quad\times
\frac{\chi(z;z_o)}{4\pi D_L^2(z;z_o)}
\tilde{\Lambda}_B\left[z;L_m(z;z_o,\dot{q}_m),L_M\right]
\,dz\,.
\end{aligned}
\label{eq:QdIV}
\end{equation}

The heating from {\em all} gamma-ray blazars is obtained simply by
setting $\dot{q}_m=0$.  On the other hand, we may set $\dot{q}_m$
implicitly via
\begin{equation}
\dot{Q}(z_o;\dot{q}_m) = \mathcal{Q} \, \dot{Q}(z_o;0)\,,
\end{equation}
where $\mathcal{Q}$ ranges from 0 to 1, yielding a heating rate
limit at each redshift that we shall call
$\dot{q}_{\mathcal{Q}}$.  From this, we may then obtain a number of
contributing blazars:
\begin{multline}
{\mathcal N}_{B,IV}(z_o;\mathcal{Q})
=
\int_{z_o}^{\infty} 4\pi D_A^2(z;z_o) \frac{dD_P}{dz}\\
\times
\tilde{\Phi}_B\left[z;L_m(z;z_o,\dot{q}_{\mathcal{Q}}),L_M\right] \,dz\,.
\label{eq:NIV}
\end{multline}
This is shown for $\mathcal{Q}=0.5$ and $\mathcal{Q}=0.75$ in Figure \ref{f:NBs}
by the black solid and long-dashed lines,
respectively.  While both are similar to the other measures of ${\mathcal N}_B$
below $z\sim1$,
 above this redshift they fall much more rapidly.  This is
due to the shift of $\tBLF(z,L)$ towards higher luminosities, and thus
the luminosity density (and therefore heating rate) becomes dominated
by fewer, more luminous sources.  Nevertheless, ${\mathcal N}_B$ is a rapidly
increasing function of $\mathcal{Q}$, as evidence by the fact that
${\mathcal N}_{B,IV}$ increases by approximately an order of magnitude when
$\mathcal{Q}$ increases from $0.5$ to $0.75$.

\subsection{Location and Properties of the Sources Responsible for the Bulk of the Heating}
Armed with a definition of ${\mathcal N}_B$, we can now address which class of
high-energy gamma-ray blazars dominates the heating rate.  That is, we
can assess whether the local heating is dominated by close,
intrinsically dim objects or by distant, intrinsically luminous
sources.
This is done simply by inspecting the integrands in Equations
(\ref{eq:QdIV}) and (\ref{eq:NIV}).
However, to interpret these, we will first build some
intuition based upon an extremely simplified model, for which an
analytical result is trivially obtained.

\subsubsection{Static Euclidean Universe}

\begin{figure}
\begin{center}
\includegraphics[width=0.95\columnwidth]{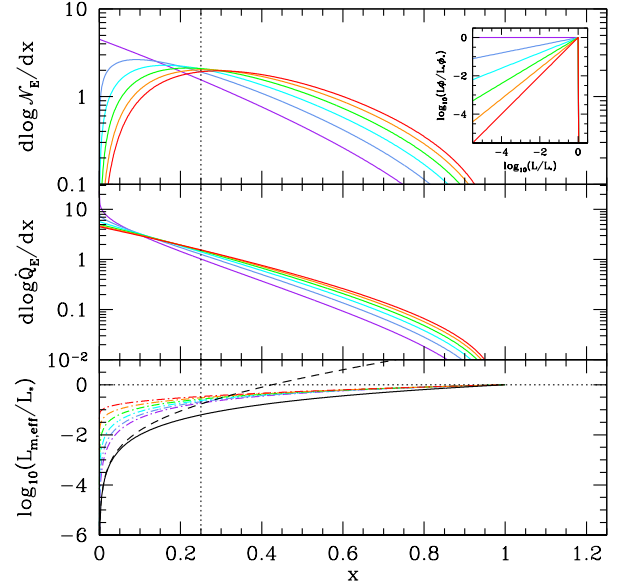}
\end{center}
\caption{
  Radial distribution ($x\equiv D/D_*$) of the sources that contribute
  to the total 
  number (top) and heating rate (middle), and the
  limiting and effective luminosity (bottom) for a flux-limited sample
  of sources 
  described by a fixed, cutoff power-law luminosity function in a
  static Euclidean Universe.
  Different colors correspond to different luminosity function
  power-laws ($\xi=-1$, $-0.8$, $-0.6$, $-0.4$, $-0.2$, and $0$ shown
  in violet, blue, cyan, green, orange, and red), with the
  associated luminosity-weighted luminosity 
  function shown explicitly in the inset.
  For reference, the radius at which $D=\Dpp$ is shown by the vertical
  dotted line.   In comparison to the approximate $L_m$ we used (black
  solid), we show the luminosity limit when absorption is included by
  the black dashed line, and the $L_{\rm eff}$ defined in Equation
  (\ref{eq:LeffE}) are shown by the dot-dash lines.
  This may be compared directly with Figure \ref{f:NBds}.
}
\label{f:EUds}
\end{figure}

\begin{figure}
\begin{center}
\includegraphics[width=0.95\columnwidth]{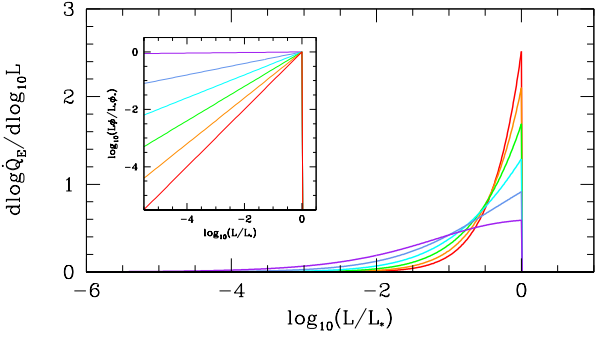}
\end{center}
\caption{$d\log\dot{Q}_E/d\log_{10}L$ for a flux-limited sample of
  sources described by a fixed, cutoff power-law luminosity function
  in a static Euclidean Universe.  Different 
  colors correspond to different luminosity function power-laws, with
  the associated luminosity-weighted luminosity function shown
  explicitly in the inset.  This may be compared directly with Figure
  \ref{f:dQdL}.
}
\label{f:dQEdL}
\end{figure}

We begin with a toy problem in which we consider a fixed power-law
luminosity function in a static, Euclidean Universe.
Specifically, we choose,
\begin{equation}
\tilde{\phi}(L) = \tilde{\phi}_* \left(\frac{L}{L_*}\right)^\xi \Theta\left(\frac{L_*}{L}\right)\,,
\end{equation}
where $\Theta(x)$ is the Heaviside function, $\tilde{\phi}_*$ is the overall
normalization of the luminosity function, $L_*$ is a maximum
luminosity (approximating a break), and $\xi$ is an arbitrary
constant.   We will also assume that $\Dpp$ is independent of
energy, i.e., there is some characteristic value for which
$\tau=D/\Dpp$ and we may bring it out of the energy integral in the
definition of $\dot{q}$.  As a consequence, a fixed limit in
$\dot{q}_m$ corresponds directly to a fixed flux limit, $F_m$.

In a Euclidean Universe the number of objects takes the particularly
simple form,
\begin{equation}
\begin{aligned}
{\mathcal N}_E
&=
\int_0^\infty dD 4\pi D^2 e^{-D/\Dpp} \int_{\log_{10}L_m}^\infty \tilde{\phi}(L) \,\dlL\\
&=
\int_0^\infty dD 4\pi D^2 e^{-D/\Dpp}\, \Theta\left(\frac{L_*}{L_m}\right) \\
&\qquad\qquad\qquad\times\int_{\log_{10}L_m}^{\log_{10}L_*} \tilde{\phi}_* \left(\frac{L}{L_*}\right)^\xi \,\dlL\,.
\end{aligned}
\end{equation}
The flux limit, $F_m$, gives a luminosity limit of $L_m=4\pi D^2 F_m$,
where we have ignored the optical depth.  Since we will be most
concerned with how peaked the various integrands are at nearby
distances, this is not a significant oversight (including it would
serve to make them only more so).  The definition of $L_m$ implies
a maximum distance as well, with $D_*=\sqrt{L_*/4\pi F_m}$, and
therefore $L_m/L_* = (D/D_*)^2\equiv x^2$.  Thus, we have
\begin{equation}
\begin{aligned}
{\mathcal N}_E
&=
\frac{4\pi\tilde{\phi}_* }{\ln10} \int_0^{D_*} dD D^2 e^{-D/\Dpp} \xi^{-1}\left[ 1 -  \left(\frac{L_m}{L_*}\right)^\xi \right]\\
&=
\frac{4\pi D_*^3 \tilde{\phi}_* }{\ln10} \int_0^1 dx 
\,\xi^{-1} x^2 \left( 1 - x^{2\xi} \right) e^{-x/x_{\rm pp}}\,.
\end{aligned}
\end{equation}
From this we trivially obtain
\begin{equation}
\frac{d{\mathcal N}_E}{dD}
=
\frac{1}{D_*} \frac{d{\mathcal N}_E}{dx}
=
\frac{4\pi D_*^2 \tilde{\phi}_* }{\ln10}
\xi^{-1} x^2 \left(1-x^{2\xi}\right)  e^{-x/x_{\rm pp}}\,,
\end{equation}
providing some notion of the location of the most numerous sources.
Typical values of $\xi$ range from $-1$ to $0$, in practice, and
$d\log{\mathcal N}_E/dx$ is shown in the top panel of Figure \ref{f:EUds} for
a variety of choices of $\xi$ within this range. 

The integral for $\dot{Q}$ is similarly simple,
\begin{equation}
\begin{aligned}
\dot{Q}_E
&=
\int_0^\infty dD 4\pi D^2 e^{-D/\Dpp}
\int_{\log_{10}L_m}^\infty \frac{L}{4\pi D^2 \Dpp} \tilde{\phi}(L) \,\dlL\\
&=
\frac{\tilde{\phi}_* L_* D_*}{\ln10 \Dpp} \int_0^{1} dx\,
\frac{1-x^{2\xi+2}}{1+\xi}\, e^{-x/x_{\rm pp}}\,,
\end{aligned}
\end{equation}
and thus,
\begin{equation}
\frac{d\dot{Q}_E}{dD}
=
\frac{\tilde{\phi}_* L_*}{\ln10 \Dpp}\,
\frac{1-x^{2\xi+2}}{1+\xi}\, e^{-x/x_{\rm pp}}\,,
\end{equation}
giving an idea of the location of the sources responsible for the bulk
of the heating.
For a variety of $\xi$, $d\log\dot{Q}_E/dx$ is shown in the middle
panel of Figure \ref{f:EUds}.

Generally, we find that for all but the largest $\xi$, 
$d\log {\mathcal N}_E/dx$
and $d\log\dot{Q}_E/dx$ are peaked at small distances.  In particular,
both are typically dominated by $x<x_{\rm pp}$.  The bottom panel of
Figure \ref{f:EUds} shows the flux-limited $L_m$ with (dashed) and
without (solid) absorption included.  Including absorption suppress
the contributions at large $x/x_{\rm pp}$, forcing 
$d\log {\mathcal N}_E/dx$ and $d\log\dot{Q}_E/dx$ to be even more
strongly peaked at small distances.

Since $L_m$ is a strong function of $D$, contributions from different
distances have different luminosity distributions.  It is possible to
roughly characterize this by defining a typical luminosity, 
$L_{\rm eff}$, associated with contributions at a given $D$:
\begin{equation}
L_{\rm eff}
\equiv
4\pi D^2 \Dpp \frac{d\dot{Q}_E/dD}{d{\mathcal N}_E/dD}
=
L_* \frac{\xi}{1+\xi}
\frac{1-x^{2\xi+2}}{1-x^{2\xi}}
\,.
\label{eq:LeffE}
\end{equation}
With $L_m$, this is shown in the bottom panel of Figure \ref{f:EUds}.
Generally $L_{\rm eff}$ is larger than $L_m$, and for small $x$
substantially so.  Thus, even for nearly flat $\tilde{\phi}(L)$
($\xi\sim-1$) the objects that contribute to the heating are {\em not}
dominated by the numerous, intrinsically dim objects with luminosities
$L\sim L_m$.

Alternatively, we may perform the integral over $D$ first, in which
the flux limit implies a maximum distance to which a given object can
be seen, $D_M=\sqrt{L/4\pi F_m}$, providing some insight into
luminosity of the sources responsible for the heating.  Doing so
yields
\begin{equation}
\begin{aligned}
\frac{d\dot{Q}_E}{d\log_{10}L}
&=
\frac{D_*}{\Dpp}
L\tilde{\phi}(L)
\int_0^{D_M/D_*} e^{-x/x_{\rm pp}} dx\\
&=
\frac{D_*}{\Dpp}
L\tilde{\phi}(L)
\left(1-e^{-\sqrt{L/L_*}/x_{\rm pp}}\right)\,.
\end{aligned}
\end{equation}
This is shown in Figure \ref{f:dQEdL} for a variety of $\xi$.  In all
cases the luminosity at the break in $\tilde{\phi}(L)$, i.e., $L_*$,
contributes most significantly to the heating rate.  However, the
relative importance of lower-luminosity objects does depend upon the
luminosity function; flat luminosity functions (i.e., $\xi=-1$) have
many low-luminosity sources, and thus induce more broad
$d\log\dot{Q}_E/d\log_{10}L$. 
Nevertheless, it appears that the heating rate in our simple toy model
is generally dominated by nearby objects near the peak in the
luminosity function.

\subsubsection{TeV Blazars in the Standard Cosmology} \label{sec:RTBs}

\begin{figure}
\begin{center}
\includegraphics[width=0.95\columnwidth]{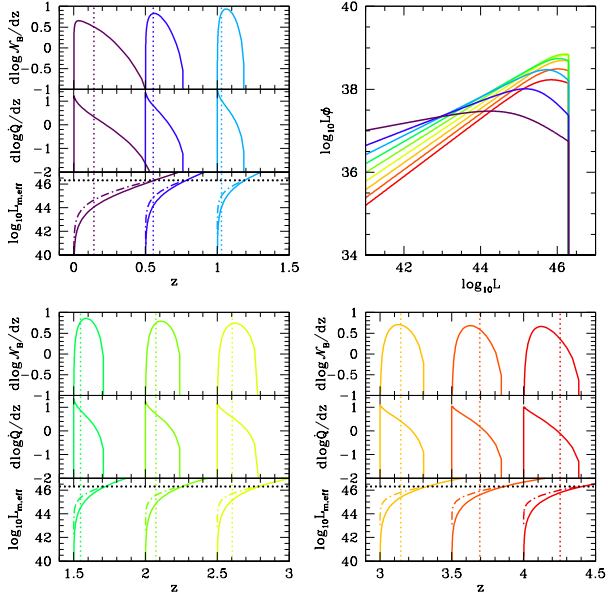}
\end{center}
\caption{
  Redshift distribution of the sources that contribute to the total
  number (top-subpanel) and heating rate (middle-subpanel), and the
  limiting luminosity (bottom-subpanel) for a fractional
  $\dot{\mathcal{Q}}$-limited sample of objects.  Different colors
  correspond to different $z_o$, indicated by the left-most value of
  $z$ for which each intersects the horizontal axis ($z_o$ ranging
  from 0 to 4), and are consistent with those used in Figure
  \ref{f:dQdL}.  For reference, the redshifts at which 
  $\bar{\tau}(z;z_o)=1$ and $L_M$ are shown by the vertical and
  horizontal dotted lines, respectively.  In comparison to $L_m$, we
  also show $L_{\rm eff}$ (as defined by Equation (\ref{eq:LeffB})) by
  the dash-dot lines.  Finally, the upper-right panel shows
  $\BLF(z_o,L)$ for each of the redshifts for which distributions are
  shown in the other panels, with corresponding colors (note that this
  shows the same relative dynamic range as that in the inset of
  Figures \ref{f:EUds} and may thus be directly compared).
}
\label{f:NBds}
\end{figure}

We now return to the physically relevant case: heating due to TeV
blazars with an evolving luminosity function in an evolving
Universe.  Here we specify the distributions of the sources
responsible for producing a fraction $\mathcal{Q}$ of the total
heating rate.  In this case, we may immediately read off 
$d{\mathcal N}_B/dz$ and $d\dot{Q}/dz$, the analogs of 
$d{\mathcal N}_E/dD$ and $d\dot{Q}_E/dD$, from 
Equations (\ref{eq:NIV}) and (\ref{eq:QdIV}), respectively, yielding,
\begin{equation}
\begin{gathered}
\frac{d{\mathcal N}_B}{dz}(z;z_o) = 4\pi D_A^2(z;z_o) \frac{dD_P}{dz}
\tilde{\Phi}_B\left[z;L_m(z;z_o,\dot{q}_{\mathcal{Q}}),L_M\right]\\
\begin{aligned}
\frac{d\dot{Q}}{dz}(z;z_o)
&=
4\pi D_A^2(z;z_o) \frac{dD_P}{dz}
\left(\frac{1+z}{1+z_o}\right)^{2-\alpha}\\
&~~~\,\quad\times\frac{\chi(z;z_o)}{4\pi D_L^2(z;z_o)}
\tilde{\Lambda}_B\left[z;L_m(z;z_o,\dot{q}_{\mathcal{Q}}),L_M\right]\,.
\end{aligned}
\end{gathered}
\end{equation}
These are shown, normalized by their integrated values, for $z_o$
ranging from $0$--$4$ in Figure \ref{f:NBds} for $\mathcal{Q}=0.5$.
In addition we show an analogously defined characteristic luminosity,
\begin{equation}
\begin{aligned}
L_{\rm eff}(z;z_o)
&\equiv
\frac{4\pi D_L^2(z;z_o)}{\chi(z;z_o)} \frac{d\dot{Q}/dz}{d{\mathcal N}_B/dz}\\
&=
\left(\frac{1+z}{1+z_o}\right)^{2-\alpha}
\frac{
  \tilde{\Lambda}_B\left[z;L_m(z;z_o,\dot{q}_{\mathcal{Q}}),L_M\right]
}{
  \tilde{\Phi}_B\left[z;L_m(z;z_o,\dot{q}_{\mathcal{Q}}),L_M\right]
}\,.
\end{aligned}
\label{eq:LeffB}
\end{equation}

The generic features of our static toy model are also apparent here.
At all observer redshifts the heating rate is dominated by the nearest
sources.  At low $z_o$, where $L\tBLF$ is nearly flat, the number of
objects is also heavily weighted towards nearby objects, well within
the redshift at which $\bar{\tau}=1$.  However, $d{\mathcal N}_B/dz$ and
$d\dot{Q}/dz$ evolve with observer redshift due to both, the intrinsic
evolution of $\tBLF$ and the background Universe.  As a consequence, by
$z_o\sim0.5$ the peak of $d{\mathcal N}_B/dz$ has moved to the $\bar{\tau}=1$
redshift, implying that $z_1$ (where $\bar{\tau}(z_1,z_o)=1$) is not a particularly accurate estimate
of the redshifts that contribute significantly to the heating.
This is further supported by the high-$z_o$ behavior of $d{\mathcal N}_B/dz$,
which once again is heavily weighted at redshifts inside of
$z_1$ due to onset of the decline in the blazar population.

The typical luminosities of objects responsible for the heating also
evolve.  At $z_o\sim0$ these are roughly $10^{44}\,\erg\,\s^{-1}$,
rising to $3\times10^{45}\,\erg\,\s^{-1}$ by $z_o\sim2$.  We also
compute the heating rate per logarithmic decade in luminosity:
\begin{multline}
\frac{d\dot{Q}}{d\log_{10}L}
=
\int_{z_o}^{z_m} 4\pi D_A^2(z;z_o) \frac{dD_P}{dz}
\left(\frac{1+z}{1+z_o}\right)^{2-\alpha}\\
\times
\frac{\chi(z;z_o)}{4\pi D_L^2(z;z_o)}
L\tBLF(z,L)\,,
\end{multline}
(where $z_m$ is determined implicitly by
$L_m(z_m;z_o,\dot{q}_{\mathcal{Q}})=L$) shown in Figure \ref{f:dQdL}
for a number of $z_o$.
At $z_o\sim0$ the distribution of heating rates is a relatively broad
function of $L$ centered near $10^{45}\,\erg\,\s$.  Until $z_o\sim2$,
as $z_o$ grows $d\dot{Q}/d\log_{10}L$ becomes increasingly peaked and
centered upon increasingly larger luminosities.  Above $z_o\sim2$ this
trend reverses, though the distribution of luminosities that contribute
appreciably to the heating rate never becomes comparable to that in
the present epoch.  Thus, generally it appears that the heating is due
predominantly to nearby objects with luminosities comparable to
$10^{45}\,\erg\,\s^{-1}$.

\end{appendix}

\bibliographystyle{apj}

\end{document}